\documentclass[amsmath,aps,prb,twocolumn,longbibliography,superscriptaddress,floatfix]{revtex4-2}
\usepackage{graphicx}
\usepackage{amsmath}
\usepackage{amssymb}
\usepackage{bm}
\usepackage{enumerate}
\usepackage{xfrac}
\usepackage[caption=false]{subfig}
\usepackage[normalem]{ulem}
\usepackage[usenames,dvipsnames]{xcolor}
\usepackage{txfonts}
\usepackage[mathscr]{euscript}
\usepackage[pdftex,
pdflang={en-US},
colorlinks=true,
urlcolor=blue,
citecolor=blue,
pdfauthor={Pontus Laurell},
pdftitle={{Prediction of superconductivity by hole doping the Haldane spin-1 chain}}]{hyperref}
\usepackage{verbatim}
\usepackage{ulem}
\usepackage{subdepth}

\usepackage[stretch=10,shrink=10]{microtype}

\makeatletter
\def\maketitle{
	\@author@finish
	\title@column\titleblock@produce
	\suppressfloats[t]}
\makeatother

\begin{document}

\title{Luther-Emery liquid and dominant singlet superconductivity in the hole-doped Haldane spin-1 chain}
\author{Pontus Laurell}
\affiliation{Department of Physics and Astronomy, University of Tennessee, Knoxville, Tennessee 37996, USA.}
\author{Jacek Herbrych}
\affiliation{Institute of Theoretical Physics, Faculty of Fundamental Problems of Technology, Wroc\l{}aw University of Science and Technology, 50-370 Wroc\l{}aw, Poland}
\author{Gonzalo Alvarez}
\affiliation{Computational Sciences and Engineering Division, Oak Ridge National Laboratory, Oak Ridge, Tennessee 37831, USA}
\author{Elbio Dagotto}
\affiliation{Department of Physics and Astronomy, University of Tennessee, Knoxville, Tennessee 37996, USA.}
\affiliation{Materials Science and Technology Division, Oak Ridge National Laboratory, Oak Ridge, Tennessee 37831, USA}

\date{\today}

\begin{abstract}
	We investigate the pairing tendencies in the hole-doped Haldane spin-1 chain. To allow for doping, we extend the original spin chain Hamiltonian 
	into a fermionic model involving a  two-orbital Hubbard chain at intermediate or strong repulsive interaction strengths $U$, and for degenerate orbitals. At half-filling and large $U$, the ferromagnetic Hund's coupling, $J_\mathrm{H}$, generates effective spin-$1$ moments, with antiferromagnetic correlations between sites. 
	Using large-scale density matrix renormalization group calculations, we study accurately the system's behavior under light hole-doping. 
	For $U=1.6$ in units of the non-interacting bandwidth and for $J_\mathrm{H}/U\gtrsim 0.275$ we find that singlet pairing dominates the long-distance physics, establishing this system as a promising platform for repulsively mediated superconductivity. We provide concrete
examples of materials that could realize the physics described here. 
	We also provide evidence that the system approaches a Luther-Emery liquid state at large system sizes, 
	reminiscent of the behavior of doped one-orbital two-leg ladders at weak coupling, which also have superconducting tendencies. 
	The numerically calculated central charge approaches one in the thermodynamic limit, indicating a single gapless mode as is expected for the Luther-Emery state. Exponents characterizing the power-law decays of singlet pair-pair and charge density-density correlations are determined, and found to approximately satisfy the Luther-Emery identity.
\end{abstract}
\maketitle

\section{Introduction}

Doped spin-$1/2$ Mott insulators have received considerable attention as a route to high-$T_c$ superconductivity in, e.g., the cuprate superconductors \cite{RevModPhys.66.763, RevModPhys.78.17}. The two-dimensional $t-J$ and one-band Hubbard models \cite{Arovas2022} are often proposed as minimal models in this context. Their solution have proven a longstanding challenge, but have seen significant recent progress due to advances in numerical techniques and computing power \cite{Jiang2021, doi:10.1146/annurev-conmatphys-090921-033948}. A more tractable version of this problem occurs in quasi-one-dimensional geometries---including chains and ladders---which are well-suited to research 
by numerically exact approaches such as the density matrix renormalization group (DMRG) \cite{PhysRevLett.69.2863, PhysRevB.48.10345}. Remarkably, these geometries are also experimentally relevant \cite{Elbio1992, Elbio1996, Dagotto_1999, Vuletic2006, RMP2013, Zhang2017, Ying2017, Tseng_2022, Scheie2024Cuprate, Padma2024Cuprate} to, e.g., cuprate and iron-based ladder materials---several of which exhibit pressure-induced superconductivity \cite{doi:10.1143/JPSJ.65.2764, PhysRevB.72.064512, Takahashi2015, Yamauchi2015}---as well as supramolecular crystals \cite{doi:10.1021/acs.nanolett.1c03236} and quantum simulation using ultracold atoms \cite{Hirthe2023}.

Another enticing approach is to dope spin-$1$ Mott insulators \cite{doi:10.1126/science.289.5478.419, PhysRevResearch.2.023184, PhysRevB.106.045103}, and in particular the much-studied Haldane spin-$1$ chain \cite{PhysRevLett.50.1153, PhysRevLett.59.799, Affleck1989a}, which has symmetry-protected topological states \cite{PhysRevB.80.155131, PhysRevB.81.064439, PhysRevB.85.075125} and non-local order parameters  \cite{PhysRevB.40.4709, Kennedy1992}.  
Haldane spin chain physics emerges naturally at strong coupling in systems where the low-energy physics can be captured by a two-orbital Hubbard model with repulsive electron-electron interactions, and where the ferromagnetic Hund's coupling $J_\mathrm{H}$ is strong enough to favor locally aligned spins. Most work in this context has focused on simplified models such as two-leg spin-$1/2$ $t-J$ \cite{PhysRevLett.79.713, doi:10.1143/JPSJ.69.1946, doi:10.1143/JPSJ.70.547, Jiang2018, PhysRevB.99.094510} and Hubbard ladders \cite{SHIRAKAWA2007663, NISHIMOTO20071059, PhysRevB.77.224510} with ferromagnetic rung couplings to generate effective $S=1$ moments on each rung. Both bosonization \cite{PhysRevB.52.6189, PhysRevB.53.14036, doi:10.1143/JPSJ.65.2241} and numerical studies \cite{doi:10.1143/JPSJ.69.1946, doi:10.1143/JPSJ.70.547} indicate that such models have finite spin gaps and pairing tendencies (hole pair formation), which is robust to perturbations affecting the two orbitals equally.

\begin{figure}[t]
	\includegraphics[width=\columnwidth]{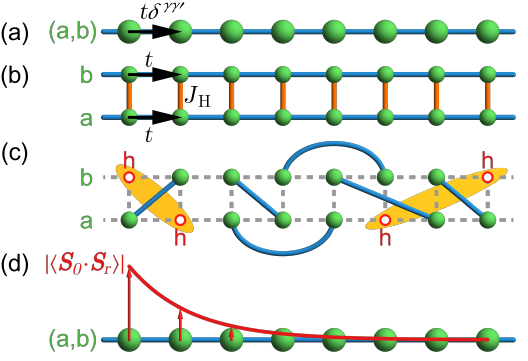}
	\caption{\label{fig:overview}
		Overview. (a) The two-orbital Hubbard chain considered in this work. Each site hosts two orbitals, labeled a and b (green text). 
		(b) Alternative representation as a two-leg ladder, where each site hosts a single orbital. Interorbital interactions such as Hund's coupling are represented as rung couplings in orange. 
		(c) A representative component of the doped orbital resonating valence bond state, 
		which at half-filling was shown to provide a good representation of the Haldane spin-1 chain ground state for the AKLT model \cite{Patel2020}. Here spin-1/2 singlets (blue lines) are of range longer than nearest-neighbors to adapt to the Haldane state, with a spin correlation length longer than in the AKLT model.  Upon introduction of several holes (red circles), hole pairs are formed as indicated by the shaded orange regions. The remaining electrons are in a superposition of states with singlets over various distances, in all possible combinations with equal weight. 
		(d) Spin correlation in the chain. The ferromagnetic Hund's coupling favors a net spin at each site, which becomes a robust spin 1 at large $U$, while antiferromagnetic correlations between sites are generated by electron-electron interactions. The exponential decay is due to the spin gap in the system.
	}
\end{figure}
\begin{figure*}
	\includegraphics[width=\textwidth]{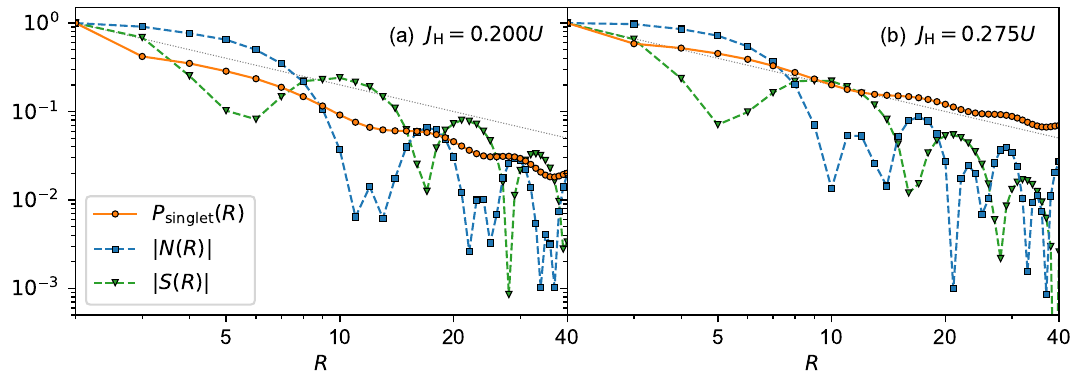}
	\caption{\label{fig:corrs}
		Comparison of correlation functions. The decays of the normalized singlet pair-pair $P_\mathrm{singlet}(R)$, density-density $N(R)$, and spin-spin correlations $S(R)$ with distance $R$ are contrasted for (a) $J_\mathrm{H}=0.200U$ and (b) $J_\mathrm{H}=0.275U$. Both panels are for $U/W=1.6$ and $L=96$ at $x=1/12$ hole doping. The dotted lines indicate a power law decay with exponent $\alpha=1$. The two panels showcase the trend where the singlet pair-pair correlations become dominant and long range (i.e. decaying slower than $R^{-1}$) at high $J_\mathrm{H}/U$ values. An expanded version of this figure showing the evolution for additional values of $J_\mathrm{H}/U$ is provided in Appendix~\ref{app:corrs}.
	}
\end{figure*}

More recently, orbitally degenerate two-orbital Hubbard-Kanamori chains with full inter- and intraorbital electron-electron interactions have been found to display qualitative tendencies towards 
spin-singlet hole pair formation at intermediate Hubbard repulsion \cite{PhysRevB.96.024520, Patel2020}. In the following we will refer to this system, illustrated in Figure~\ref{fig:overview}, as the two-orbital Hubbard chain (TOHC). 
It is a remarkable system, with obvious deep connections with the paradigmatic Haldane spin chain. 
At half-filling, its entanglement spectrum \cite{PhysRevLett.101.010504} and string order parameter suggest a transition from a topologically trivial state at $U=0$ to the Haldane phase at relatively weak $U$ \cite{Patel2020, Jazdzewska2023}. 
The presence of edge states was also demonstrated \cite{Jazdzewska2023}, highlighting that our system is a rare example of a correlated topological state. 
In addition, an orbital resonating valence-bond (ORVB) state was introduced to explain the precursors of singlet superconductivity in the system \cite{Patel2020}. This state is a linear superposition of the Affleck-Kennedy-Lieb-Tasaki (AKLT) valence-bond states familiar from the generalized spin-1 chain problem including biquadratic terms \cite{PhysRevLett.59.799}, and provides a liquid background of preformed singlets, as compared to the more rigid background
of rung singlets in the two-leg ladders. Upon hole doping, effective singlet hole pairing is expected; see Fig.~\ref{fig:overview}(c). Intuitively this occurs because upon hole doping the system tries to minimize 
the number of preformed spin-1/2 singlets which are broken by doping, thus
effectively inducing the binding of pairs of holes.
Extending the analogy to Haldane chain physics even further, it was found that an easy-plane anisotropy term can drive the system into a topologically trivial triplet pairing regime \cite{Jiang2018, Patel2020}. These prior results strongly suggest, but do not prove, that superconductivity indeed dominates.

In this work, we study the TOHC in detail, reporting results for significantly larger system sizes than previously studied. 
In Patel \emph{et al.} \cite{Patel2020} superconductivity precursors such as pair formation were identified. 
Here, via large-scale DMRG calculations we show that these pairs form a quantum coherent state. Specifically, for the first time we
find that the singlet pair-pair correlation becomes dominant for $J_\mathrm{H}/U\gtrsim 0.275$, indicating that \emph{the TOHC is a promising platform for repulsively mediated superconductivity.} 
Notably, this occurs in a system that combines electronic correlation effects with nontrivial topology, since a nonzero Hubbard repulsion is required to generate the superconductivity, and the Haldane chain has protected spin-$1/2$ edge states. 
Moreover, we show that the system approaches a Luther-Emery-like state \cite{PhysRevLett.33.589} in the thermodynamic limit, with one gapless charge mode and a spin gap. This is reflected in the central charge, which tends to one for large systems. 
We also confirm that the Luther-Emery identity for the power law decays of the singlet pair-pair and density-density correlations is approximately satisfied. In addition, we propose concrete 
materials that may realize the physics discussed here upon doping. 
We encourage the experimental study of the specific materials proposed herein to test our predictions.

\section{Model}
We consider the Hamiltonian $H=H_0+H_I$, where the non-interacting term is given by
\begin{align}
	H_0	&=	\sum_{j,\sigma,\gamma,\gamma^\prime} t^{\gamma\gamma^\prime} \left( c_{j,\gamma,\sigma}^\dagger c^{\vphantom\dagger}_{j+1,\gamma^\prime,\sigma} + \mathrm{H.c.}\right),
\end{align}
and $c_{j\gamma\sigma}$ annihilates an electron with orbital index $\gamma$ and spin projection $\sigma$ at site $j$ of the chain. $\mathrm{H.c.}$ denotes Hermitian conjugate. The hopping matrix $t^{\gamma\gamma^\prime}=t\delta^{\gamma\gamma^\prime}$ is spin-conserving and, for simplicity, diagonal in orbital space, resulting in a noninteracting bandwidth $W=4|t|$. We use $|t|=1$ as the energy unit throughout this paper.

The interaction part is of the standard Hubbard-Kanamori type,
\begin{align}
	H_I	&=	U\sum_{i,\gamma} n_{i\gamma\uparrow}n_{i\gamma\downarrow} 	+ \left( U^\prime - \frac{J_\mathrm{H}}{2}\right) \sum_{i,\gamma<\gamma^\prime} n_{i\gamma}n_{i\gamma^\prime}\nonumber\\
	&-	2J_\mathrm{H} \sum_{i,\gamma<\gamma^\prime} \mathbf{S}_{i\gamma} \cdot \mathbf{S}_{i\gamma^\prime} + J_\mathrm{H} \sum_{i,\gamma<\gamma^\prime} \left( P_{i\gamma}^\dagger P_{i\gamma^\prime} + \mathrm{H.c.}\right),	\label{eq:ham:interact}
\end{align}
where $n_{i\gamma\sigma}=c^\dagger_{i\gamma\sigma} c^{\vphantom\dagger}_{i\gamma\sigma}$ is the number operator, $U>0$ is the intra-orbital Hubbard repulsion, and the second term describes inter-orbital density-density interactions. $J_\mathrm{H}$ represents the Hund's coupling strength. 
We assume the standard relation $U'=U-2J_H$, which arises due to spin-rotational invariance.
Physically, it is expected that $U'>J_\mathrm{H}$ \cite{PhysRevB.82.104508, Dai2012}, which holds for $J_\mathrm{H}/U<1/3$. 
In this paper we report results for $0.2\leq J_\mathrm{H}/U\leq 0.35$, where the value $0.35$ is included to show that the results do not change drastically at the boundary value $1/3$. 
The third term is the Hund's coupling term, and the fourth is the on-site inter-orbital electron-pair hopping with $P_{i\gamma^\prime}=c_{i\gamma^\prime\uparrow}c_{i\gamma^\prime\downarrow}$. 
The spin-$1/2$ operators in Eq. \eqref{eq:ham:interact} are defined as $S_{i\gamma}^\alpha	=	\frac{1}{2}\sum_{\sigma\sigma'} c_{i\gamma\sigma}^\dagger \tau_{\sigma\sigma'}^\alpha  c^{\vphantom\dagger}_{i\gamma\sigma'},$
where $\alpha\in\{x,y,z\}$ and $\vec{\tau}=\left( \sigma^x,\sigma^y,\sigma^z \right)$ is the vector of Pauli matrices.

\section{Methods}

\subsection{Numerical technique}
We study ground state properties of our model with zero-temperature DMRG \cite{PhysRevLett.69.2863, PhysRevB.48.10345}, using the DMRG++ software \cite{Alvarez2009}. We work with finite systems and open boundary conditions. The system can either be represented as a length-$L$ chain with a two-orbital basis on each site, or as a length-$L$ two-leg ladder with one orbital on each site and a total of $2L$ sites; see Fig.~\ref{fig:overview}(a),(b). Although the two representations are mathematically equivalent, the ladder representation was found to perform  better, and was thus used throughout this work. 

Care was taken to achieve the best convergence possible within the memory available to us (up to $1000$ GiB). 
Using up to $m=11\,000$ DMRG states, we obtained truncation errors below $10^{-7}$ for the majority of sizes ($L\leq 192$) and $J_\mathrm{H}$ values, and below $10^{-6}$ for the rest (only affecting $J_\mathrm{H}/U\leq 0.25$). In general, convergence was easier at higher $J_\mathrm{H}/U$, while full entanglement scaling at the lowest $J_\mathrm{H}/U$ was not always possible. 
Explicit reorthogonalization was used to avoid Lanczos ghost states. Further details on how to reproduce the numerical results are provided in the Supplemental Material \cite{Supplemental}.

\subsection{Correlation functions}
We define the general singlet pair creation operator as in Ref.~\cite{Patel2020}
\begin{align}
	\Delta_{(i,j){-}}^{\gamma\gamma'\dagger}	&=	\frac{1}{\sqrt{2}} \left[ c_{i\gamma\uparrow}^\dagger c_{j\gamma'\downarrow}^\dagger - c_{i\gamma\downarrow}^\dagger c_{j\gamma'\uparrow}^\dagger \right],
\end{align}
from which pair-pair correlation functions are constructed. We focus on nearest-neighbor singlet pairs odd under orbital exchange, which has previously been established as the dominant pairing channel for the parameters we study \cite{PhysRevB.96.024520, Patel2020}. We also consider on-site interorbital triplet pairs in Appendix~\ref{app:corrs}. The singlet pair creation operator is given by
\begin{align}
	\Delta_{\mathrm{S},\mathrm{nn}}^\dagger(i)	&=	\Delta^{ab\dagger}_{(i,i+1)-} - \Delta^{ba\dagger}_{(i,i+1)-},
\end{align}
from which the singlet pair-pair correlations are defined as
\begin{align}
	P_\mathrm{singlet}(R)	&= \frac{1}{N_R} \sum_i \left\langle \Delta_{\mathrm{S},\mathrm{nn}}^\dagger (i) \Delta_{\mathrm{S},\mathrm{nn}}^{\vphantom\dagger} (i+R) \right\rangle,	\label{eq:pairpair:singlet}
\end{align}
%%%% This is 1/N_R in our convention, because the pair creation operators \Delta are normalized by 1/sqrt(2).
%%%% If \Delta is not normalized, the sum in P should have 1/(2N_R) as in \cite{PhysRevB.96.024520}
where $N_R$ denotes the number of total neighbors at distance $R$ from site $i$, summed over all sites. 
We also define the spin-spin and density-density correlation functions
\begin{align}
	S(R)	&=	\frac{1}{N_R} \left[ \sum_i 
	\langle S_i^z S_{i+R}^z\rangle - \langle S_i^z\rangle \langle S_{i+R}^z\rangle \right]	\\
	&=\frac{1}{3N_R} \left[ \sum_i 
	\langle \mathbf{S}_i \cdot \mathbf{S}_{i+R}\rangle - \langle \mathbf{S}_i\rangle\cdot \langle \mathbf{S}_{i+R}\rangle \right],\\
	N(R)	&=	\frac{1}{N_R} \left[ \sum_i 
	\langle n_i n_{i+R}\rangle - \langle n_i\rangle \langle n_{i+R}\rangle \right].
\end{align}
In calculating these correlation functions we neglect one quarter of the chain at each end to avoid edge effects. The correlation functions are then normalized to their values at distance $R=2$ to enable comparing the relative decay rates.

\section{Results}
\subsection{Dominant singlet superconductivity}

\begin{figure}
	\includegraphics[width=\columnwidth]{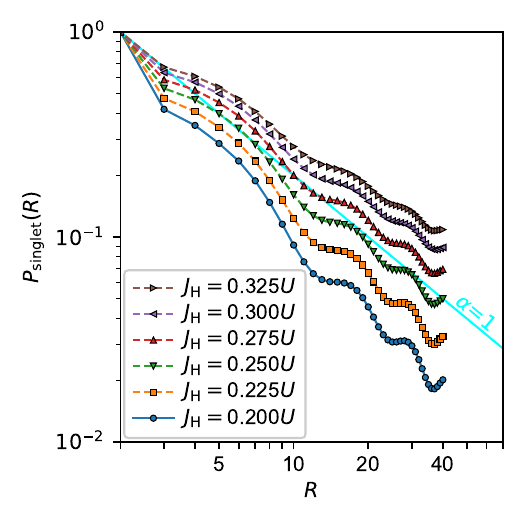}
	\caption{\label{fig:corrs:JH}
		The singlet pair-pair correlations at different $J_\mathrm{H}/U$ ratios are compared. The correlations become long range, i.e. decay slower than $R^{-1}$ (indicated by the cyan line), for $J_\mathrm{H}/U\gtrsim 0.275$.}
\end{figure}
Previous studies of correlation functions in the TOHC were limited to chains of length $L=48$ \cite{PhysRevB.96.024520, Patel2020}, primarily due to memory constraints. Here we report results for chain lengths up to $L=96$, allowing for cleaner analysis of the long-distance behavior, and, more importantly, for the precise determination of exponents characterizing the decay of correlation functions with distance. This information is crucial to determine the universality class of the ground state. See the Methods for details about the numerical method. We focus on the case of weak hole doping, with hole density 
$x=\frac{n}{2L}=1/12$, 
where $n$ is the number of holes, and choose $U/W=1.6$. (Half-filling corresponds to $x=0$.) According to the previously studied phase diagram for $J_\mathrm{H}/U=0.25$ \cite{Patel2020} using smaller systems, these parameters correspond to a phase where singlet superconductivity is qualitatively expected to dominate.

Figure~\ref{fig:corrs} compares normalized pair-pair, spin-spin, and density-density correlations for $J_\mathrm{H}/U=0.2$ and $J_\mathrm{H}/U=0.275$. The definitions of these correlation functions are provided in the Methods. It is clear that, for $J_\mathrm{H}/U=0.275$ [Fig.~\ref{fig:corrs}(b)], the singlet pair-pair correlations decay slower than $R^{-1}$ and thus dominate at long distance. In contrast, at $J_\mathrm{H}/U=0.2$ [Fig.~\ref{fig:corrs}(a)] the singlet pair-pair correlations decay more rapidly, and the density-density and spin-spin correlations become more important. This dependence of the singlet pair-pair correlations on $J_\mathrm{H}/U$ is illustrated more directly in Fig.~\ref{fig:corrs:JH}, where we see a crossover between $J_\mathrm{H}/U=0.25$ and $J_\mathrm{H}/U=0.275$. The trend of increasingly fast decay as $J_\mathrm{H}/U$ is decreased is expected to continue if $J_\mathrm{H}/U$ is lowered further, compatible with the binding energy results of Refs.~\cite{PhysRevB.96.024520, Patel2020} that suggest pairing will no longer occur at small $J_\mathrm{H}/U$. At $J_\mathrm{H}/U=0.25$ the singlet pair-pair correlations decay as $R^{-\alpha}$, with ${\alpha}\approx 1.04$ determined by a power-law fit. The same exponent for $J_\mathrm{H}/U=0.275$ is $\alpha \approx0.92$. As we will discuss later, if the system is in a Luther-Emery liquid state, an exponent $\alpha>1$ indicates a phase dominated by charge density-density correlations, whereas $\alpha<1$ indicates a superconducting phase.

We note that the singlet pair-pair correlation remains positive at all $R$, whereas the density-density correlations oscillate across zero, leading to spikes in $|N(R)|$ in Fig.~\ref{fig:corrs}, c.f. Ref.~\cite{PhysRevB.92.195139}. The spin-spin correlations $S(R)$ also oscillate across zero, stemming from the parent antiferromagnetic state at half-filling. These short-range oscillations are invisible in Fig.~\ref{fig:corrs} as $|S(R)|$ is plotted. The visible longer-range oscillations 
are caused by finite-size effects that, fortunately, do not affect the pair-pair correlations of our main focus. In fact, these pair-pair correlations behave very smoothly with increasing $R$.

\subsection{Energy gaps and entanglement}
\begin{figure}
	\includegraphics[width=\columnwidth]{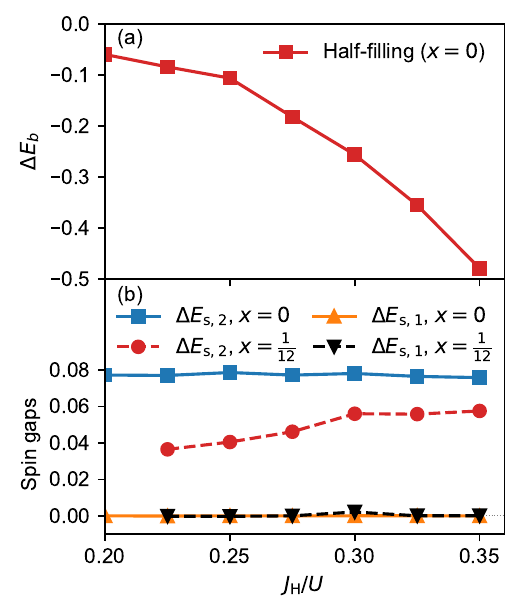}
	\caption{\label{fig:gaps}Finite-size-scaled energy gaps. In both panels, gaps are given in units of the hopping energy $|t|$.
		(a) The binding energy at $U/W=1.6$ and half-filling depends strongly on the value of $J_\mathrm{H}/U$. For the dependence on $U/W$, see Ref.~\cite{Patel2020}.  
		(b) Spin gaps at $U/W=1.6$ and half-filling (solid lines), and at a hole doping concentration of $x=1/12$ (dashed lines). The conventional spin gap $\Delta E_\mathrm{s}(1,x)$ is zero throughout the $J_\mathrm{H}/U$ range as expected for the half-filled TOHC with open boundary conditions \cite{Jazdzewska2023}. Physically, this arises because of the connection with the Haldane spin chain at $U\gg W$, which features a ground state degeneracy linked to the formation of $S=1/2$ edge states \cite{Kennedy_1990, PhysRevB.52.12844}. As seen here, the effect is present also in the doped TOHC. Thus, the physical spin excitation gap is instead given by $\Delta E_\mathrm{s}(2,x)$ representing $\Delta S=2$ excitations. The latter gap is found to remain open. It is remarkably flat at half-filling within the range of $J_\mathrm{H}/U$ considered here, but is known to vary substantially for lower $J_\mathrm{H}/U$ and for lower $U/W$ values \cite{Jazdzewska2023}. At finite doping it decreases as $J_\mathrm{H}/U$ is lowered in the studied range, unlike in the half-filled case. This effect may be understood as a promotion of the kinetic energy by the dopants, which for weak doping is expected to modify the spin gap similarly to how it is modified at half-filling by reducing $U/W$. 
		Due to the challenging convergence at finite doping and magnetization, we have not obtained spin gaps for the doped system $J_\mathrm{H}/U=0.2$.
	}
\end{figure}
We next consider the energy gaps in the system. The binding energy $\Delta E_\mathrm{b}$ at half-filling is shown as a function of $J_\mathrm{H}/U$ in Fig.~\ref{fig:gaps}(a). It is defined as \cite{Patel2020, PhysRevB.96.024520, RevModPhys.66.763}
\begin{align}
	\Delta E_\mathrm{b}	&=	E(2) - E(0) - 2\left[ E(1)-E(0)\right] = e_2 - 2e_1,
\end{align}
where $E(n)$ is ground state energy for $n$ holes (relative to half-filling) and $e_n=E(n)-E(0)$ denotes the energy of the $n$-hole state, measured relative to the undoped case. The subscript $\mathrm{b}$ denotes binding. 
When negative,  $\Delta E_\mathrm{b}$ signals the presence of a two-hole bound state; a necessary condition for pairing to occur similar to Cooper pair formation. The results indicate that the bound state potential well becomes deeper as $J_\mathrm{H}/U$ increases, in agreement with the increasingly strong pair-pair correlations. We also consider the spin gaps
\begin{align}
	\Delta E_{\mathrm{s}} \left(\Delta S^z, x\right)	&=	E(\Delta S^z, x) - E(0, x),
\end{align}
where $E(\Delta S^z, x)$ denotes the energy in the $\Delta S^z$ magnetization sector for hole density $x$. The subscript $\mathrm{s}$ is used to denote spin gap. Due to the similarities to the Haldane spin chain, we expect that $\Delta E_{\mathrm{s}}(1,x)$ vanishes due to the presence of spin-$1/2$ edge states, and that the physical spin gap is instead given by $\Delta E_{\mathrm{s}}(2,x)$  \cite{Kennedy_1990, PhysRevB.52.12844, Jazdzewska2023}. The spin gaps at half-filling and finite doping are shown in Fig.~\ref{fig:gaps}(b). The physical spin gap, corresponding to $\Delta S=2$ excitations, is finite at all $J_\mathrm{H}/U$ values considered. The system-size dependence is shown in the Supplemental Material \cite{Supplemental}.

\begin{figure}
	\includegraphics[width=\columnwidth]{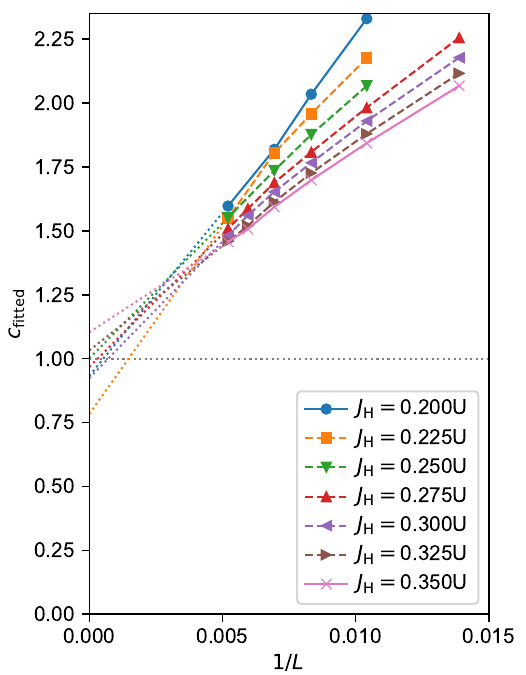}
	\caption{\label{fig:entanglement}Central charge. The fitted central charge $c$ as a function of $1/L$ and $J_\mathrm{H}/U$. 
	The dotted lines are linear interpolations of the plotted data. The results are consistent with each $c_{J_\mathrm{H}}\rightarrow 1$ in the thermodynamic limit, with deviations caused by numerical errors. The behavior is overall consistent with a C$1$S$0$ state and Luther-Emery physics.}
\end{figure}
Three signs point towards the possibility of a Luther-Emery liquid state in the hole-doped TOHC: i) there is a finite spin gap $\Delta E_\mathrm{s}(2,x)$, ii) long-range singlet pair-pair correlations are observed at large Hund's coupling, and iii) the TOHC is formally similar to the two-leg one-band Hubbard ladder at weak coupling, which  
is considered an archetypal Luther-Emery system. Indeed, the two orbitals can be represented as fictious legs in a two-leg ladder [Fig.~\ref{fig:overview}(b)]. 

An additional criterion for the Luther-Emery liquid state is that there is a single gapless charge mode, producing a so-called C1S0 state (in this notation, a C$m$S$n$ state has $m$ gapless charge modes and $n$ gapless spin modes.) To investigate this mode we study the entanglement entropy. 
Although strong finite-size effects are noted at low $L$, the trends stabilize for $L\geq 96$; see Appendix~\ref{app:entanglement} for details. 
Here we extracted the central charge at fixed system size by fitting the entropy to the conformal field theory prediction \cite{Calabrese2004}
\begin{align}
	S(j)	&=	\frac{c}{6}	\ln \left[ \frac{L}{\pi}\sin\left( \frac{\pi j}{L}\right)\right] + C,
\end{align}
where $C$ is a non-universal constant, and $j$ is the position along the chain.  
The results are shown in Fig.~\ref{fig:entanglement}. By inspection, it is clear that the central charge for each $J_\mathrm{H}/U$ is approaching $1$. Interpreting the central charge as the number of gapless modes and recalling the presence of a spin gap, the results point towards a C1S0 state.

We noticed that the fitted central charge depends strongly on the system size, producing unusually high entanglement for low system sizes. This has the curious consequence that the DMRG truncation error at fixed bond dimension can be {\it smaller} for chain lengths $L\gtrsim 96$ than for short and intermediate system sizes. A similarly strongly size-dependent behavior of the central charge was observed in the one-orbital Hubbard two-leg ladder at weak $U$ \cite{PhysRevB.102.115136}. That system features an initial renormalization group flow towards a perturbatively unstable C$2$S$1$ fixed line, before eventually tending to a C$1$S$0$ Luther-Emery state. It is unclear whether a similar picture holds for the TOHC, however a related renormalization group analysis at weak coupling finds a C${1}$S$0$ state \cite{PhysRevB.85.115406}. For symmetry-breaking hopping matrices, namely including nonzero off-diagonal components and different diagonal hoppings for each orbital, the phase diagram may be more complex, with a number of gapless modes that depends on $J_\mathrm{H}/U$ \cite{PhysRevB.98.184517}. Studies of the range of stability of the Luther-Emery liquid state in the two-orbital model
when using generic hopping matrices and crystal fields will be computer-time demanding and it is postponed 
for future work.

\begin{figure*}
	\includegraphics[width=\textwidth]{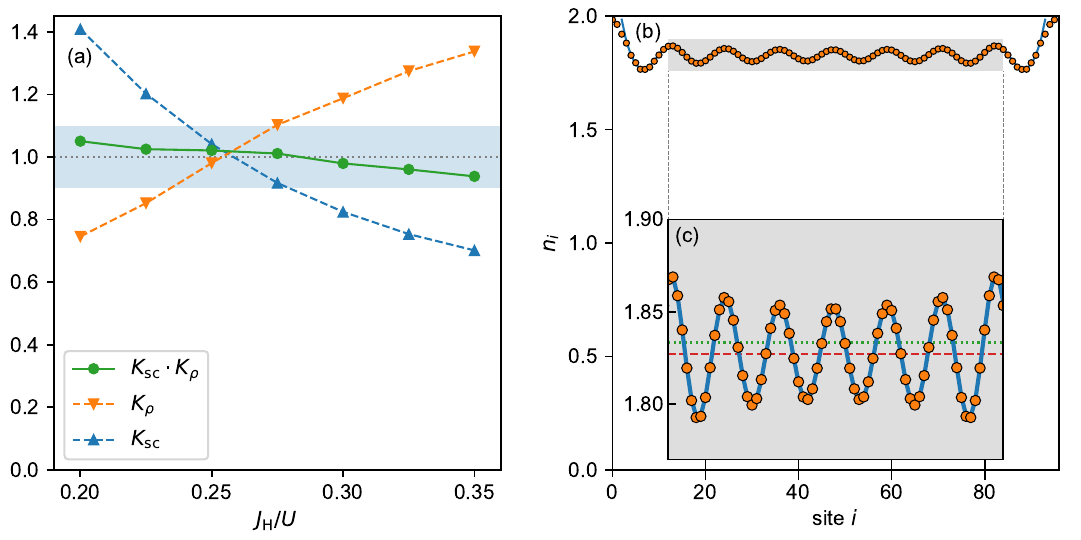}
	\caption{\label{fig:exponents}Luttinger exponents. (a) Test of the Luther-Emery identity. The product $K_\mathrm{sc}\cdot K_\rho$ (green circles) remains close to $1$ throughout the studied range of $J_\mathrm{H}/U$, consistent with a Luther-Emery state. The shaded region represents a band of $\pm 10\%$. Also shown are the scaling exponents $K_\mathrm{sc}$ from singlet pair-pair correlations, and $K_\rho$ extracted from the charge densities. The shown data is for $L=96$, $U/W=1.6$, $x=1/12$ hole doping, and open boundary conditions. 
	(b) Example fit of the local charge density profile at $J_\mathrm{H}=0.275U$, $L=96$ and $x=1/12$ hole doping. To avoid the divergent boundary effects, the fit is performed only for data in the shaded region.
	The inset (c) provides a zoomed-in view of the shaded region, emphasizing the Friedel oscillations induced by the open boundaries. The dashed red line indicates the fitted density offset $n_0$, and the dotted green line indicates the average filling $\langle n\rangle$ for reference. 
	}
\end{figure*}
\subsection{Luttinger exponents}
In one-dimensional systems, the long-distance decays of the singlet pair-pair and charge density-density correlations are generally expected to follow power laws
\begin{alignat}{1}
	P_\mathrm{singlet}(R)	&\propto	R^{-K_\mathrm{sc}},	\label{eq:singlet:powerlaw}\\
	{\left| N(R)\right|}					&\propto	R^{-K_\rho},
\end{alignat}
up to modulations periodic in $R$ and higher-order corrections. 
In the Luther-Emery state the exponents satisfy the identity $K_\mathrm{sc} \cdot K_\rho = 1$ \cite{PhysRevB.92.195139, PhysRevB.53.R2959}. In practice,  numerical results on ladders at weak and intermediate coupling often deviate from this identity due to the  challenging convergence properties of correlation functions \cite{NOACK1996281, PhysRevB.56.7162, PhysRevB.92.195139, Shen2023}. 

In our case, Figures~\ref{fig:corrs} and \ref{fig:corrs:JH} show that $P_\mathrm{singlet}$ exhibits clear power law behavior with minimal oscillations, and we extract $K_\mathrm{sc}$ by direct fitting. In contrast, $N(R)$ features pronounced oscillations. To avoid modeling the modulation, $K_\rho$ was instead obtained by fitting Friedel oscillations in the local charge density (induced by the open boundaries) \cite{PhysRevB.92.195139, PhysRevB.65.165122} to 
\begin{equation}
\langle n_j\rangle  =  \delta n \frac{\cos \left( \pi N_\mathrm{h} j/L_\mathrm{eff} + \phi_1 \right)}{\left[ L_\mathrm{eff} \sin \left( \pi j/L_\mathrm{eff} + \phi_2\right) \right]^{K_\rho/2}} + n_0,
\end{equation}
where $n_j=\sum_\gamma n_{j,\gamma}$ is the density operator on site $j$ (summed over orbitals $\gamma$), $\delta n$ is a non-universal amplitude, $n_0$ is the background density, $\phi_1$ and $\phi_2$ are phase shifts, $N_\mathrm{h}$ is the number of holes in the system, and $L_\mathrm{eff}\lesssim L_x$ is an effective length that is shorter than $L$ due to the finite extent of the hole pairs. We treat all six variables (i.e. $\delta_n$, $n_0$, $\phi_1$, $\phi_2$, $K_\rho$ and $L_\mathrm{eff}$) as fitting parameters, obtaining $L-4 \lesssim L_\mathrm{eff}\lesssim L-3$. An example fit is shown in Fig.~\ref{fig:exponents}(b-c).

The Luttinger exponents $K_\mathrm{sc}$ (extracted from the singlet pair-pair correlations shown in Figs.~\ref{fig:corrs}, \ref{fig:corrs:JH} by fitting to power laws) and $K_\rho$ (extracted from density oscillations) are shown in Fig.~\ref{fig:exponents}(a) along with their product. The product is close to $1$ for all studied values of $J_\mathrm{H}/U$, consistent with Luther-Emery liquidity. There is a clear crossover from dominant density-density correlations ($K_\rho<1$, $K_\mathrm{sc}>1$) at low $J_\mathrm{H}/U$ to dominant singlet pair-pair correlations ($K_\mathrm{sc}<1$, $K_\rho>1$) at high $J_\mathrm{H}/U$.

\section{Conclusion} In this publication, we show that upon hole doping an electronic generalization of Haldane's spin-1 model, 
the system becomes a superconductor. 
While previous work suggested this conclusion via the convincing proof of Cooper pair formation, the large scale density matrix renormalization group study reported here allows us to finally computationally conclude that the model is indeed dominated by singlet pairing in a range of couplings and after hole doping. 
At large system sizes we find that the TOHC behaves as a Luther-Emery liquid, clarifying the nature of the system in the thermodynamic limit. This finding highlights the role of universality classes in determining the long-distance physics even for realistic multiorbital models with many competing energy scales. 
We have also demonstrated that the TOHC features dominant singlet superconductivity for $J_\mathrm{H}/U\geq 0.275$, with a crossover into the long-range superconducting phase likely occurring in the range $0.25 < J_\mathrm{H}/U <0.275$. Although slightly higher than the value $J_\mathrm{H}/U=0.25$ often used for iron-based superconductors \cite{PhysRevB.82.104508, Dai2012}, such Hund's coupling strengths are physical, and may be found in other multiorbital compounds. It should be noted that phase transitions in Hund-correlated quantum matter often depend on the interplay between the Hund's coupling and the Hubbard interactions. Whether the long-range superconducting phase can be stabilized at lower $J_\mathrm{H}/U$ by tuning $U/W$ or by introducing further-range hopping processes or by using a nearest-neighbor hopping matrix different from the unit matrix is left for future work.

To realize this physics in materials, two nearly degenerate orbitals are required
\footnote{\textcolor{red}{Our preliminary results on short chains with fixed $J_H/U=0.25$ suggest that i) the binding and pair-pair correlations are robust to the introduction of interorbital hopping, although the optimal $U/W$ ratio changes, ii) the binding and pair-pair correlations survive up to at least 10\% detuning of the diagonal intraorbital hopping, and iii) the physics is sensitive to crystal field effects. Indeed, a crystal field raising the energy of one orbital by $0.03W$ already inhibits binding in $L=16$ chains. Thus deviations from a diagonal nearest-neighbor hopping matrix may be acceptable, but care should be taken to find materials with minimal crystal field splitting.}}. This rules out many compounds already known to realize Haldane spin chain physics, such as the nickel-based Y$_2$BaNiO$_5$ \cite{PhysRevB.54.R6827}, which has significant level splitting and may be in an entirely different regime \cite{PhysRevB.106.045103}. Nevertheless, quasi-one-dimensional materials with two highly degenerate orbitals are certainly possible, as evidenced by materials such as OsCl$_4$ \cite{10.1063/5.0079570}. However, its $U/W$ ratio may be too high to be relevant for our superconducting mechanism at intermediate coupling, instead justifying a spin-1 chain description \cite{PhysRevB.101.155112, PhysRevB.105.014435}. The currently leading candidates are compounds like RuOCl$_2$ and OsOCl$_2$ \cite{PhysRevB.105.174410, 10.1063/5.0023729}, which have $U\approx W$ and $J_\mathrm{H}/U=0.2$. 
These strongly anisotropic van der Waals materials also feature subleading interchain hoppings within the plane and very weak interplane hoppings \cite{PhysRevB.105.174410}. When the purely one-dimensional superconducting state discussed in this article is dominant, such interchain couplings may stablize it into a true long-range order. 
Further research into candidate materials and experimental realizations of
the ideas presented in this publication should be pursued.

Access to the computational results reported in this paper will be made available from Ref. \cite{data}.

\begin{acknowledgments}
We thank N. Kaushal, L.-F. Lin, B. Pandey and Y. Zhang for helpful discussions. 
The work of PL and ED was supported by the U.S. Department of Energy (DOE), Office of Science, Basic Energy Sciences (BES), Materials Sciences and Engineering Division. The work of GA was supported by the U.S. Department of Energy, 	Office of Science, National Quantum Information Science Research Centers, Quantum Science Center. 
\end{acknowledgments}

\appendix
\section{Additional correlation function results}\label{app:corrs}
\begin{figure*}
	\includegraphics{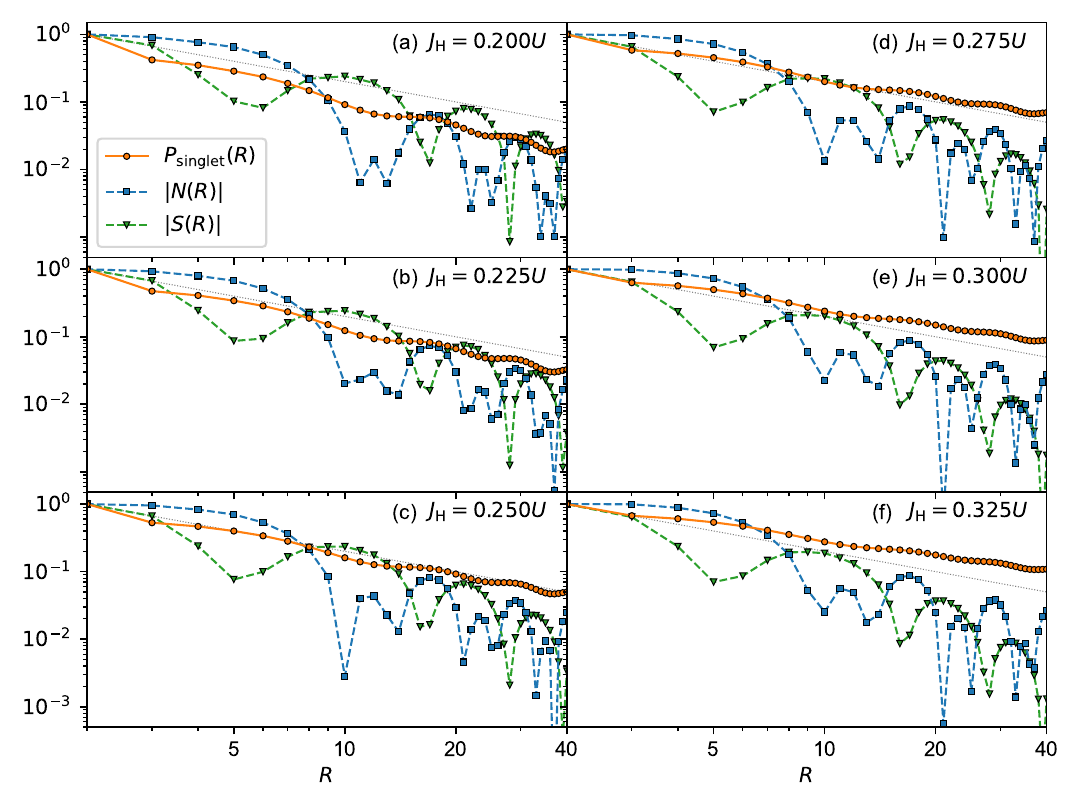}
	\caption{\label{fig:corrs:extended}
		Comparison of correlation functions. The decays of the normalized singlet pair-pair, density-density, and spin-spin correlations are contrasted for (a) $J_H=0.200U$, (b) $J_H=0.225U$, (c) $J_H=0.250U$, (d) $J_H=0.275U$, (e) $J_H=0.300U$, and (f) $J_H=0.325U$.  
		All panels are for $U/W=1.6$ and $L=96$ at $x=1/12$ hole doping. The dotted lines indicate a power law decay with exponent $\alpha=1$.
		There is a clear trend towards the singlet pair-pair correlations becoming dominant and long range (i.e. decaying slower than $R^{-1}$) at high $J_\mathrm{H}/U$ values.
	}
\end{figure*}

Fig.~\ref{fig:corrs:extended} shows the comparison of normalized singlet pair-pair, charge density-density, and spin-spin correlations for a range of $J_\mathrm{H}$ values. By comparing the singlet pair-pair correlations (orange) to the dotted line indicating $R^{-1}$ decay, it is clear that there is a crossover from the region $J_\mathrm{H}/U\geq 0.275$, where $P_\mathrm{singlet}$ decays slower than $R^{-1}$, to the low $J_\mathrm{H}/U$ region, where the decay is faster than $R^{-1}$.

\begin{figure}
	\includegraphics{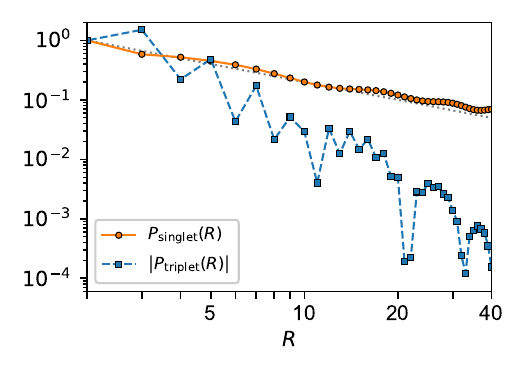}
	\caption{\label{fig:triplet}Triplet pair-pair correlations at $J_\mathrm{H}/U=0.275$. The triplet pair-pair and the singlet pair-pair correlations are contrasted for $U/W=1.6$, $J_\mathrm{H}/U=0.275$, and $x=1/12$ for a system of length $L=96$. The dotted line indicates $R^{-1}$.}
\end{figure}

In the main text we only discussed the $P_\mathrm{singlet}$ correlations (defined in the Methods). Following Refs.~\cite{PhysRevB.96.024520, Patel2020} we also considered on-site interorbital triplet pairs, for which the correlation function is defined
\begin{align}
	P_\mathrm{triplet}(R)	&=	\frac{1}{N_R} \sum_i \left\langle T_\mathrm{on}^\dagger (i) T_\mathrm{on} (i+R) \right\rangle,	\label{eq:pairpair:triplet}
\end{align}
where
\begin{align}
	T_\mathrm{on}^\dagger(i)	=	\Delta^{ab\dagger}_{(i,i)+}&=\frac{1}{\sqrt{2}} \left[ c_{ia\uparrow}^\dagger c_{ib\downarrow}^\dagger + c_{ia\downarrow}^\dagger c_{ib\uparrow}^\dagger \right],
\end{align}
and the general inter-site triplet pair creation operator is given by
\begin{align} 
	\Delta_{(i,j)+}^{\gamma\gamma'\dagger}	&=	\frac{1}{\sqrt{2}} \left[ c_{i\gamma\uparrow}^\dagger c_{j\gamma'\downarrow}^\dagger + c_{i\gamma\downarrow}^\dagger c_{j\gamma'\uparrow}^\dagger \right].
\end{align}
Although the triplet pair-pair correlations can be stabilized by an easy axis anisotropy \cite{Patel2020}, they are exponentially suppressed for the case considered here, with vanishing easy axis anisotropy. This is exemplified in Fig.~\ref{fig:triplet} for $J_\mathrm{H}/U=0.275$.

\section{Additional entanglement properties}\label{app:entanglement}
\begin{figure}
	\includegraphics{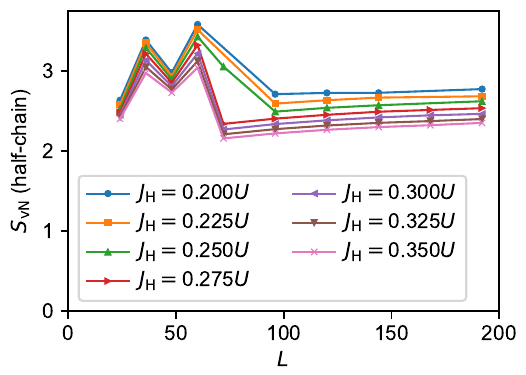}
	\caption{\label{fig:entanglement:entropy}Half-chain entanglement entropy. At small system sizes, the half-chain von Neumann entanglement entropy varies wildly with the system size. Above $L=96$ it scales approximately logarithmically for all $J_\mathrm{H}/U$ considered. The shown data is for $U/W=1.6$ and $x=1/12$.
	}
\end{figure}
The half-chain entanglement entropy is shown in Fig.~\ref{fig:entanglement:entropy}. Conformal field theory predicts that the half-chain entanglement entropy of a critical system with open boundary conditions scales logarithmically with system size, according to $S_\mathrm{vN} \propto  \frac{c}{6}	\ln L$, where $c$ is the central charge \cite{Calabrese2004}. By fitting the numerical data to this relation, we find $c\approx 1.15$--$1.21$ for $J_\mathrm{H}/U\geq 0.25$, consistent with $c\rightarrow 1$ in the thermodynamic limit. The numerical data for  $J_\mathrm{H}/U=0.2$ and $J_\mathrm{H}/U=0.225$ are associated with higher truncation errors, and are thus less reliable. The fit for $J_\mathrm{H}/U=0.225$ gives $c\approx 0.79$, which is also consistent with $c\rightarrow 1$. However, the fit for $J_\mathrm{H}/U=0.2$ gives $c\approx 0.52$. We believe this value is a result of insufficient convergence.

A striking consequence of the high entanglement entropy at small system sizes is that the required bond dimension to reach a given truncation error can be higher than it would be at large sizes. This highlights that the system-size dependence of certain quantities, such as central charges and exponents related to universality classes, can be highly nontrivial in electronic systems at intermediate coupling when using DMRG.

\newpage
\clearpage
\title{Supplemental material for ``Luther-Emery liquid and dominant singlet superconductivity in the hole-doped Haldane spin-1 chain''}
\maketitle
\onecolumngrid

\section*{Finite-size scaling}
\begin{figure}
	\includegraphics{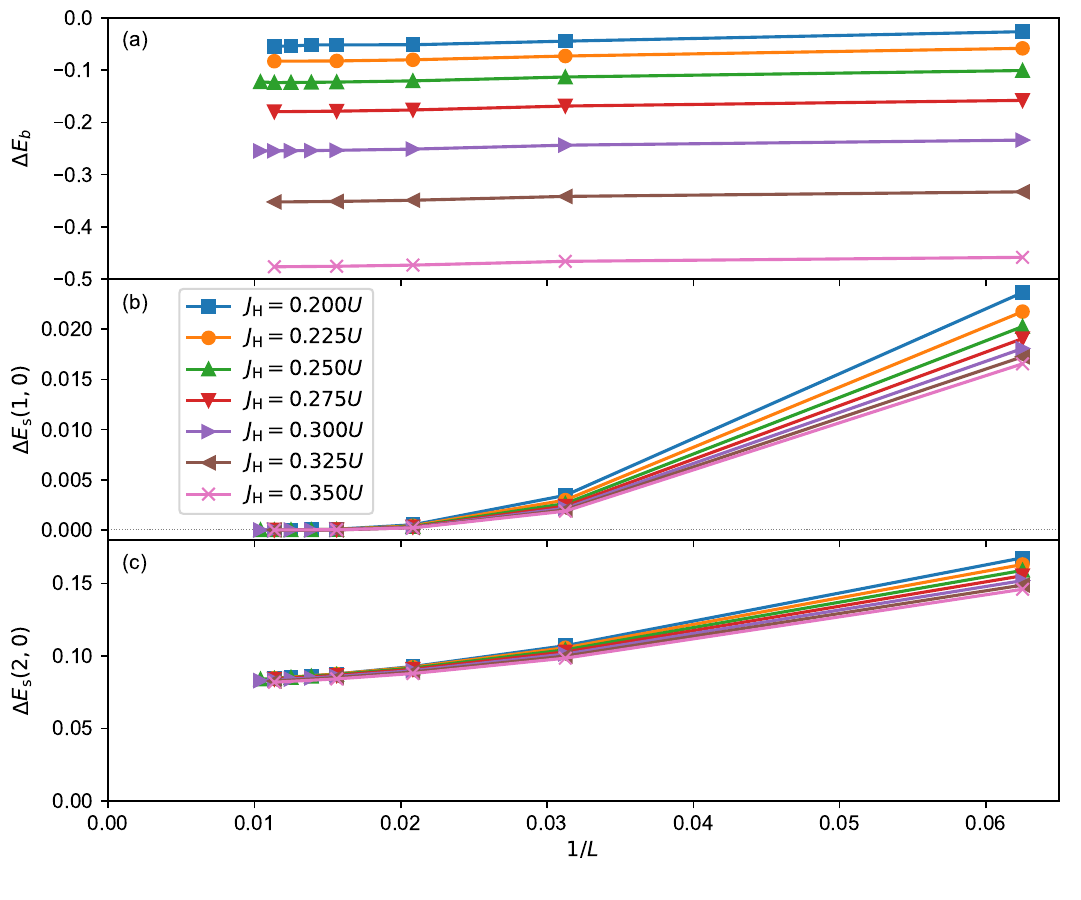}
	\caption{\label{fig:gaps:halffilling}Finite-size scaling of energy gaps at half-filling and $U/W=1.6$. 
		(a) The binding energy has only a minor size dependence, in agreement with Ref.~\cite{PhysRevB.96.024520}.
		(b) The conventional spin gap $\Delta E_\mathrm{s}(1,0)$ quickly approaches zero as $L$ is increased.
		(c) The physical spin gap $\Delta E_\mathrm{s}(2,0)$ remains open. At half-filling, the finite-size-scaled magnitude of the gap is approximately the same for all studied $J_\mathrm{H}/U$ ratios.
	}
\end{figure}

\begin{figure}
	\includegraphics{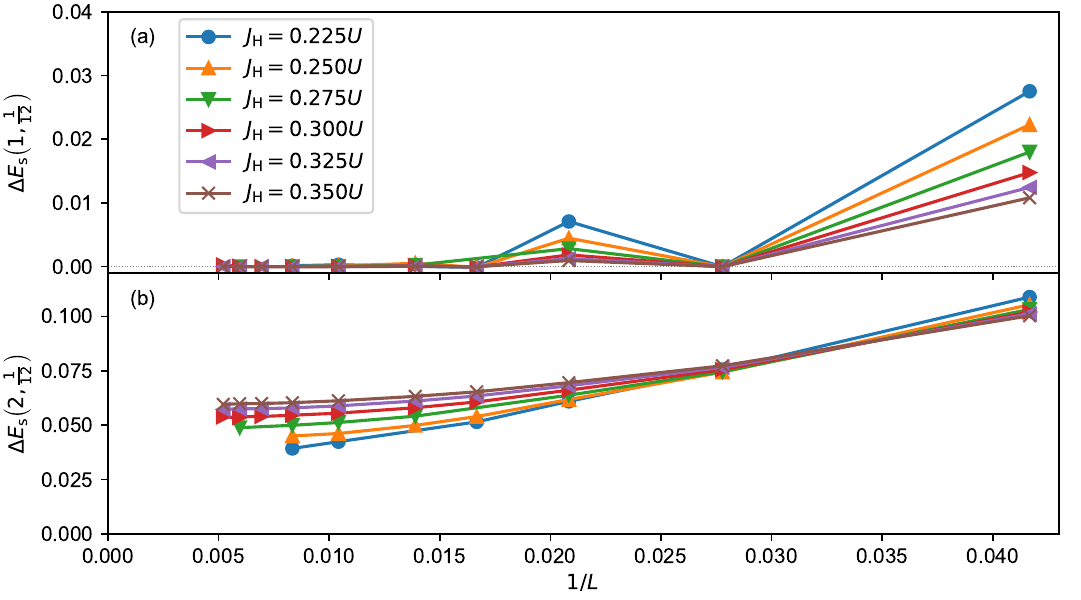}
	\caption{\label{fig:gaps:doped}Finite-size scaling of energy gaps at hole doping concentration $x=1/12$ and $U/W=1.6$. (a) The conventional spin gap $\Delta E_\mathrm{s}(1,x)$ quickly approaches zero as $L$ is increased. (b) The physical spin gap $\Delta E_\mathrm{s}(2,x)$ remains open, with a magnitude that depends on $J_\mathrm{H}/U$.
	}
\end{figure}

Details of the finite-size scaling of the energy gaps at half-filling and hole doping of $x=1/12$ are shown in Fig.~\ref{fig:gaps:halffilling} and Fig.~\ref{fig:gaps:doped}, respectively.


\section*{Reproducing the numerical calculations}
The numerical results reported in this work were obtained with DMRG++ version 6.05 and PsimagLite version 3.04. This supplementary note provides instructions for obtaining the software, and a schematic input file.

The DMRG++ and PsimagLite software can be downloaded using
\begin{verbatim}
git clone https://github.com/g1257/dmrgpp.git
git clone https://github.com/g1257/PsimagLite.git
\end{verbatim}
Dependencies include the BOOST and HDF5 libraries. To compile:
\begin{verbatim}
cd PsimagLite/lib; perl configure.pl; make
cd ../../dmrgpp/src; perl configure.pl; make
\end{verbatim}
The documentation can be found at 
\nolinkurl{https://g1257.github.io/dmrgPlusPlus/manual.html} or built locally
by doing \verb!cd dmrgpp/doc; make manual.pdf!.

For brevity in the following we run
\begin{verbatim}
export PATH="<PATH-TO-DMRG++>/src:$PATH"
\end{verbatim}
The ground state run for a sample input \texttt{input.ain} may now be executed using
\begin{verbatim}
dmrg -f input.ain
\end{verbatim}
Two- and four-point correlation functions need to measured in restart runs, using e.g.
\begin{verbatim}
observe -f input.ain "<gs|c?0'*c?0;c?1'*c?1|gs>" -p 12 > nupndown.txt
\end{verbatim}
where the apostrophe denotes Hermitian conjugate. This command measures $\langle n_{i\uparrow} n_{j\downarrow}\rangle$ for all $i<j$. Four-point correlations are specified by separating the $c$ operators using additional semicolons.

Below we show a schematic example input at half-filling for $|t|=1$, $U/W=1.6$, $J_\mathrm{H}=0.275$, open boundary conditions, and a ground state in the zero magnetization sector. It uses a simplified syntax supported from DMRG++ 6.05 and onward. Variables to be substituted by the user have been prefixed with a dollar sign (\$) in a shell script-style notation.  We use a two-leg ladder representation for the two-orbital Hubbard chain as described in the methods, such that the total number of sites is $N=2L$. In general, \verb!dir0! refers to the leg (or intraorbital) direction, while \verb!dir1! refers to the rung (or interorbital) direction. We note that the purpose of the schematic input is to illustrate the structure of the inputs; for real calculations on systems of reasonable size it is not sufficient to use only three finite loops or to use a maximum number of kept states as low as $500$.
\begin{verbatim}
##Ainur1.0
TotalNumberOfSites=$N;
NumberOfTerms=5;
Model=HubbardOneBandExtendedSuper;

SolverOptions="twositedmrg,calcAndPrintEntropies";
Version="version";

# Keep a maximum of m states, but allow truncation with tolerance and minimum states
TruncationTolerance="1e-6,100";
InfiniteLoopKeptStates=100;
FiniteLoops=[
[@auto, 100, 0],
[@auto, 250, 0],
[@auto, 500, @save]						# save output to .hd5 file that can be read by observe
];

# Set filling and magnetization sector.
TargetElectronsUp=$L;
TargetElectronsDown=$L;
# To search all symmetry sectors, use TargetElectronsTotal=$N instead.

# Tolerance for Lanczos
string LanczosOptions="reortho";		# enables explicit reorthogonalization
LanczosEps=1e-10;
int LanczosSteps=400;

hubbardU=[6.4,...x$N];
potentialV=[0,...x$(2N)];

### Hoppings.
gt0:DegreesOfFreedom=1;
gt0:GeometryKind="ladder";
gt0:LadderLeg=2;
gt0:GeometryOptions="ConstantValues";
gt0:dir0:Connectors=[-1];				# hopping along ladder leg
gt0:dir1:Connectors=[0.0];				# hopping along ladder rung

### U'-JH/2
gt1:DegreesOfFreedom=1;
gt1:GeometryKind="ladder";
gt1:LadderLeg=2;
gt1:GeometryOptions="ConstantValues";
gt1:dir0:Connectors=[0.0];
gt1:dir1:Connectors=[2.0];

### U[2] ~ -2JH. This is the SxSx+SySy part of the Hund's coupling term.
gt2:DegreesOfFreedom=1;
gt2:GeometryKind="ladder";
gt2:LadderLeg=2;
gt2:GeometryOptions="ConstantValues";
gt2:dir0:Connectors=[0.0];
gt2:dir1:Connectors=[-3.52];

### U[3] ~ -2JH. This is the SzSz part of the Hund's coupling term.
gt3:DegreesOfFreedom=1;
gt3:GeometryKind="ladder";
gt3:LadderLeg=2;
gt3:GeometryOptions="ConstantValues";
gt3:dir0:Connectors=[0.0];
gt3:dir1:Connectors=[-3.52];

### U[4] ~ JH. This is the pair-hopping term. In this DMRG++ model the sign is positive.
gt4:DegreesOfFreedom=1;
gt4:GeometryKind="ladder";
gt4:LadderLeg=2;
gt4:GeometryOptions="ConstantValues";
gt4:dir0:Connectors=[0.0];
gt4:dir1:Connectors=[1.76];
\end{verbatim}

Other fillings and magnetization sectors are specified by setting the \verb!TargetElectronsUp!/\verb!TargetElectronsDown! values. The solver option \verb!calcAndPrintEntropies! outputs entanglement entropies for each logical site in the ladder geometry. The entanglement entropies between physical sites correspond to ``rung cuts'' and are obtained by keeping every other entropy value in the DMRG++ output.


\begin{thebibliography}{74}%
	\makeatletter
	\providecommand \@ifxundefined [1]{%
		\@ifx{#1\undefined}
	}%
	\providecommand \@ifnum [1]{%
		\ifnum #1\expandafter \@firstoftwo
		\else \expandafter \@secondoftwo
		\fi
	}%
	\providecommand \@ifx [1]{%
		\ifx #1\expandafter \@firstoftwo
		\else \expandafter \@secondoftwo
		\fi
	}%
	\providecommand \natexlab [1]{#1}%
	\providecommand \enquote  [1]{``#1''}%
	\providecommand \bibnamefont  [1]{#1}%
	\providecommand \bibfnamefont [1]{#1}%
	\providecommand \citenamefont [1]{#1}%
	\providecommand \href@noop [0]{\@secondoftwo}%
	\providecommand \href [0]{\begingroup \@sanitize@url \@href}%
	\providecommand \@href[1]{\@@startlink{#1}\@@href}%
	\providecommand \@@href[1]{\endgroup#1\@@endlink}%
	\providecommand \@sanitize@url [0]{\catcode `\\12\catcode `\$12\catcode
		`\&12\catcode `\#12\catcode `\^12\catcode `\_12\catcode `\%12\relax}%
	\providecommand \@@startlink[1]{}%
	\providecommand \@@endlink[0]{}%
	\providecommand \url  [0]{\begingroup\@sanitize@url \@url }%
	\providecommand \@url [1]{\endgroup\@href {#1}{\urlprefix }}%
	\providecommand \urlprefix  [0]{URL }%
	\providecommand \Eprint [0]{\href }%
	\providecommand \doibase [0]{https://doi.org/}%
	\providecommand \selectlanguage [0]{\@gobble}%
	\providecommand \bibinfo  [0]{\@secondoftwo}%
	\providecommand \bibfield  [0]{\@secondoftwo}%
	\providecommand \translation [1]{[#1]}%
	\providecommand \BibitemOpen [0]{}%
	\providecommand \bibitemStop [0]{}%
	\providecommand \bibitemNoStop [0]{.\EOS\space}%
	\providecommand \EOS [0]{\spacefactor3000\relax}%
	\providecommand \BibitemShut  [1]{\csname bibitem#1\endcsname}%
	\let\auto@bib@innerbib\@empty
	%</preamble>
	\bibitem [{\citenamefont {Dagotto}(1994)}]{RevModPhys.66.763}%
	\BibitemOpen
	\bibfield  {author} {\bibinfo {author} {\bibfnamefont {E.}~\bibnamefont
			{Dagotto}},\ }\bibfield  {title} {\bibinfo {title} {Correlated electrons in
			high-temperature superconductors},\ }\href
	{https://doi.org/10.1103/RevModPhys.66.763} {\bibfield  {journal} {\bibinfo
			{journal} {Rev. Mod. Phys.}\ }\textbf {\bibinfo {volume} {66}},\ \bibinfo
		{pages} {763} (\bibinfo {year} {1994})}\BibitemShut {NoStop}%
	\bibitem [{\citenamefont {Lee}\ \emph {et~al.}(2006)\citenamefont {Lee},
		\citenamefont {Nagaosa},\ and\ \citenamefont {Wen}}]{RevModPhys.78.17}%
	\BibitemOpen
	\bibfield  {author} {\bibinfo {author} {\bibfnamefont {P.~A.}\ \bibnamefont
			{Lee}}, \bibinfo {author} {\bibfnamefont {N.}~\bibnamefont {Nagaosa}},\ and\
		\bibinfo {author} {\bibfnamefont {X.-G.}\ \bibnamefont {Wen}},\ }\bibfield
	{title} {\bibinfo {title} {Doping a {Mott} insulator: Physics of
			high-temperature superconductivity},\ }\href
	{https://doi.org/10.1103/RevModPhys.78.17} {\bibfield  {journal} {\bibinfo
			{journal} {Rev. Mod. Phys.}\ }\textbf {\bibinfo {volume} {78}},\ \bibinfo
		{pages} {17} (\bibinfo {year} {2006})}\BibitemShut {NoStop}%
	\bibitem [{\citenamefont {Arovas}\ \emph {et~al.}(2022)\citenamefont {Arovas},
		\citenamefont {Berg}, \citenamefont {Kivelson},\ and\ \citenamefont
		{Raghu}}]{Arovas2022}%
	\BibitemOpen
	\bibfield  {author} {\bibinfo {author} {\bibfnamefont {D.~P.}\ \bibnamefont
			{Arovas}}, \bibinfo {author} {\bibfnamefont {E.}~\bibnamefont {Berg}},
		\bibinfo {author} {\bibfnamefont {S.}~\bibnamefont {Kivelson}},\ and\
		\bibinfo {author} {\bibfnamefont {S.}~\bibnamefont {Raghu}},\ }\bibfield
	{title} {\bibinfo {title} {The {Hubbard} model},\ }\href
	{https://doi.org/https://doi.org/10.1146/annurev-conmatphys-031620-102024}
	{\bibfield  {journal} {\bibinfo  {journal} {Annu. Rev. Condens. Matter
				Phys.}\ }\textbf {\bibinfo {volume} {13}},\ \bibinfo {pages} {239} (\bibinfo
		{year} {2022})}\BibitemShut {NoStop}%
	\bibitem [{\citenamefont {Jiang}\ \emph {et~al.}(2021)\citenamefont {Jiang},
		\citenamefont {Scalapino},\ and\ \citenamefont {White}}]{Jiang2021}%
	\BibitemOpen
	\bibfield  {author} {\bibinfo {author} {\bibfnamefont {S.}~\bibnamefont
			{Jiang}}, \bibinfo {author} {\bibfnamefont {D.~J.}\ \bibnamefont
			{Scalapino}},\ and\ \bibinfo {author} {\bibfnamefont {S.~R.}\ \bibnamefont
			{White}},\ }\bibfield  {title} {\bibinfo {title} {Ground-state phase diagram
			of the t-t'-{J} model},\ }\href {https://doi.org/10.1073/pnas.2109978118}
	{\bibfield  {journal} {\bibinfo  {journal} {Proc. Natl. Acad. Sci. U.S.A.}\
		}\textbf {\bibinfo {volume} {118}},\ \bibinfo {pages} {e2109978118} (\bibinfo
		{year} {2021})}\BibitemShut {NoStop}%
	\bibitem [{\citenamefont {Qin}\ \emph {et~al.}(2022)\citenamefont {Qin},
		\citenamefont {Schäfer}, \citenamefont {Andergassen}, \citenamefont
		{Corboz},\ and\ \citenamefont
		{Gull}}]{doi:10.1146/annurev-conmatphys-090921-033948}%
	\BibitemOpen
	\bibfield  {author} {\bibinfo {author} {\bibfnamefont {M.}~\bibnamefont
			{Qin}}, \bibinfo {author} {\bibfnamefont {T.}~\bibnamefont {Schäfer}},
		\bibinfo {author} {\bibfnamefont {S.}~\bibnamefont {Andergassen}}, \bibinfo
		{author} {\bibfnamefont {P.}~\bibnamefont {Corboz}},\ and\ \bibinfo {author}
		{\bibfnamefont {E.}~\bibnamefont {Gull}},\ }\bibfield  {title} {\bibinfo
		{title} {The {Hubbard} model: A computational perspective},\ }\href
	{https://doi.org/10.1146/annurev-conmatphys-090921-033948} {\bibfield
		{journal} {\bibinfo  {journal} {Annu. Rev. Condens. Matter Phys.}\ }\textbf
		{\bibinfo {volume} {13}},\ \bibinfo {pages} {275} (\bibinfo {year}
		{2022})}\BibitemShut {NoStop}%
	\bibitem [{\citenamefont {White}(1992)}]{PhysRevLett.69.2863}%
	\BibitemOpen
	\bibfield  {author} {\bibinfo {author} {\bibfnamefont {S.~R.}\ \bibnamefont
			{White}},\ }\bibfield  {title} {\bibinfo {title} {Density matrix formulation
			for quantum renormalization groups},\ }\href
	{https://doi.org/10.1103/PhysRevLett.69.2863} {\bibfield  {journal} {\bibinfo
			{journal} {Phys. Rev. Lett.}\ }\textbf {\bibinfo {volume} {69}},\ \bibinfo
		{pages} {2863} (\bibinfo {year} {1992})}\BibitemShut {NoStop}%
	\bibitem [{\citenamefont {White}(1993)}]{PhysRevB.48.10345}%
	\BibitemOpen
	\bibfield  {author} {\bibinfo {author} {\bibfnamefont {S.~R.}\ \bibnamefont
			{White}},\ }\bibfield  {title} {\bibinfo {title} {Density-matrix algorithms
			for quantum renormalization groups},\ }\href
	{https://doi.org/10.1103/PhysRevB.48.10345} {\bibfield  {journal} {\bibinfo
			{journal} {Phys. Rev. B}\ }\textbf {\bibinfo {volume} {48}},\ \bibinfo
		{pages} {10345} (\bibinfo {year} {1993})}\BibitemShut {NoStop}%
	\bibitem [{\citenamefont {Dagotto}\ \emph {et~al.}(1992)\citenamefont
		{Dagotto}, \citenamefont {Riera},\ and\ \citenamefont
		{Scalapino}}]{Elbio1992}%
	\BibitemOpen
	\bibfield  {author} {\bibinfo {author} {\bibfnamefont {E.}~\bibnamefont
			{Dagotto}}, \bibinfo {author} {\bibfnamefont {J.}~\bibnamefont {Riera}},\
		and\ \bibinfo {author} {\bibfnamefont {D.}~\bibnamefont {Scalapino}},\
	}\bibfield  {title} {\bibinfo {title} {Superconductivity in ladders and
			coupled planes},\ }\href {https://doi.org/10.1103/PhysRevB.45.5744}
	{\bibfield  {journal} {\bibinfo  {journal} {Phys. Rev. B}\ }\textbf {\bibinfo
			{volume} {45}},\ \bibinfo {pages} {5744(R)} (\bibinfo {year}
		{1992})}\BibitemShut {NoStop}%
	\bibitem [{\citenamefont {Dagotto}\ and\ \citenamefont
		{Rice}(1996)}]{Elbio1996}%
	\BibitemOpen
	\bibfield  {author} {\bibinfo {author} {\bibfnamefont {E.}~\bibnamefont
			{Dagotto}}\ and\ \bibinfo {author} {\bibfnamefont {T.~M.}\ \bibnamefont
			{Rice}},\ }\bibfield  {title} {\bibinfo {title} {Surprises on the way from
			one- to two-dimensional quantum magnets: The ladder materials},\ }\href
	{https://doi.org/10.1126/science.271.5249.618} {\bibfield  {journal}
		{\bibinfo  {journal} {Science}\ }\textbf {\bibinfo {volume} {271}},\ \bibinfo
		{pages} {618} (\bibinfo {year} {1996})}\BibitemShut {NoStop}%
	\bibitem [{\citenamefont {Dagotto}(1999)}]{Dagotto_1999}%
	\BibitemOpen
	\bibfield  {author} {\bibinfo {author} {\bibfnamefont {E.}~\bibnamefont
			{Dagotto}},\ }\bibfield  {title} {\bibinfo {title} {Experiments on ladders
			reveal a complex interplay between a spin-gapped normal state and
			superconductivity},\ }\href {https://doi.org/10.1088/0034-4885/62/11/202}
	{\bibfield  {journal} {\bibinfo  {journal} {Rep. Prog. Phys.}\ }\textbf
		{\bibinfo {volume} {62}},\ \bibinfo {pages} {1525} (\bibinfo {year}
		{1999})}\BibitemShut {NoStop}%
	\bibitem [{\citenamefont {Vuletić}\ \emph {et~al.}(2006)\citenamefont
		{Vuletić}, \citenamefont {Korin-Hamzić}, \citenamefont {Ivek},
		\citenamefont {Tomić}, \citenamefont {Gorshunov}, \citenamefont {Dressel},\
		and\ \citenamefont {Akimitsu}}]{Vuletic2006}%
	\BibitemOpen
	\bibfield  {author} {\bibinfo {author} {\bibfnamefont {T.}~\bibnamefont
			{Vuletić}}, \bibinfo {author} {\bibfnamefont {B.}~\bibnamefont
			{Korin-Hamzić}}, \bibinfo {author} {\bibfnamefont {T.}~\bibnamefont {Ivek}},
		\bibinfo {author} {\bibfnamefont {S.}~\bibnamefont {Tomić}}, \bibinfo
		{author} {\bibfnamefont {B.}~\bibnamefont {Gorshunov}}, \bibinfo {author}
		{\bibfnamefont {M.}~\bibnamefont {Dressel}},\ and\ \bibinfo {author}
		{\bibfnamefont {J.}~\bibnamefont {Akimitsu}},\ }\bibfield  {title} {\bibinfo
		{title} {The spin-ladder and spin-chain system
			{(La,Y,Sr,Ca)$_{14}$Cu$_{24}$O$_{41}$}: Electronic phases, charge and spin
			dynamics},\ }\href
	{https://doi.org/https://doi.org/10.1016/j.physrep.2006.01.005} {\bibfield
		{journal} {\bibinfo  {journal} {Phys. Rep.}\ }\textbf {\bibinfo {volume}
			{428}},\ \bibinfo {pages} {169} (\bibinfo {year} {2006})}\BibitemShut
	{NoStop}%
	\bibitem [{\citenamefont {Dagotto}(2013)}]{RMP2013}%
	\BibitemOpen
	\bibfield  {author} {\bibinfo {author} {\bibfnamefont {E.}~\bibnamefont
			{Dagotto}},\ }\bibfield  {title} {\bibinfo {title} {Colloquium: The
			unexpected properties of alkali metal iron selenide superconductors},\ }\href
	{https://doi.org/10.1103/RevModPhys.85.849} {\bibfield  {journal} {\bibinfo
			{journal} {Rev. Mod. Phys.}\ }\textbf {\bibinfo {volume} {85}},\ \bibinfo
		{pages} {849} (\bibinfo {year} {2013})}\BibitemShut {NoStop}%
	\bibitem [{\citenamefont {Zhang}\ \emph {et~al.}(2017)\citenamefont {Zhang},
		\citenamefont {Lin}, \citenamefont {Zhang}, \citenamefont {Dagotto},\ and\
		\citenamefont {Dong}}]{Zhang2017}%
	\BibitemOpen
	\bibfield  {author} {\bibinfo {author} {\bibfnamefont {Y.}~\bibnamefont
			{Zhang}}, \bibinfo {author} {\bibfnamefont {L.}~\bibnamefont {Lin}}, \bibinfo
		{author} {\bibfnamefont {J.-J.}\ \bibnamefont {Zhang}}, \bibinfo {author}
		{\bibfnamefont {E.}~\bibnamefont {Dagotto}},\ and\ \bibinfo {author}
		{\bibfnamefont {S.}~\bibnamefont {Dong}},\ }\bibfield  {title} {\bibinfo
		{title} {Pressure-driven phase transition from antiferromagnetic
			semiconductor to nonmagnetic metal in the two-leg ladders
			{$A{\mathrm{Fe}}_{2}{X}_{3}$ ($A=\mathrm{Ba},\mathrm{K}$;
				$X=\mathrm{S},\mathrm{Se}$)}},\ }\href
	{https://doi.org/10.1103/PhysRevB.95.115154} {\bibfield  {journal} {\bibinfo
			{journal} {Phys. Rev. B}\ }\textbf {\bibinfo {volume} {95}},\ \bibinfo
		{pages} {115154} (\bibinfo {year} {2017})}\BibitemShut {NoStop}%
	\bibitem [{\citenamefont {Ying}\ \emph {et~al.}(2017)\citenamefont {Ying},
		\citenamefont {Lei}, \citenamefont {Petrovic}, \citenamefont {Xiao},\ and\
		\citenamefont {Struzhkin}}]{Ying2017}%
	\BibitemOpen
	\bibfield  {author} {\bibinfo {author} {\bibfnamefont {J.}~\bibnamefont
			{Ying}}, \bibinfo {author} {\bibfnamefont {H.}~\bibnamefont {Lei}}, \bibinfo
		{author} {\bibfnamefont {C.}~\bibnamefont {Petrovic}}, \bibinfo {author}
		{\bibfnamefont {Y.}~\bibnamefont {Xiao}},\ and\ \bibinfo {author}
		{\bibfnamefont {V.~V.}\ \bibnamefont {Struzhkin}},\ }\bibfield  {title}
	{\bibinfo {title} {Interplay of magnetism and superconductivity in the
			compressed {Fe}-ladder compound {${\mathrm{BaFe}}_{2}{\mathrm{Se}}_{3}$}},\
	}\href {https://doi.org/10.1103/PhysRevB.95.241109} {\bibfield  {journal}
		{\bibinfo  {journal} {Phys. Rev. B}\ }\textbf {\bibinfo {volume} {95}},\
		\bibinfo {pages} {241109(R)} (\bibinfo {year} {2017})}\BibitemShut {NoStop}%
	\bibitem [{\citenamefont {Tseng}\ \emph {et~al.}(2022)\citenamefont {Tseng},
		\citenamefont {Thomas}, \citenamefont {Zhang}, \citenamefont {Paris},
		\citenamefont {Puphal}, \citenamefont {Bag}, \citenamefont {Deng},
		\citenamefont {Asmara}, \citenamefont {Strocov}, \citenamefont {Singh},
		\citenamefont {Pomjakushina}, \citenamefont {Kumar}, \citenamefont {Nocera},
		\citenamefont {R{\o}nnow}, \citenamefont {Johnston},\ and\ \citenamefont
		{Schmitt}}]{Tseng_2022}%
	\BibitemOpen
	\bibfield  {author} {\bibinfo {author} {\bibfnamefont {Y.}~\bibnamefont
			{Tseng}}, \bibinfo {author} {\bibfnamefont {J.}~\bibnamefont {Thomas}},
		\bibinfo {author} {\bibfnamefont {W.}~\bibnamefont {Zhang}}, \bibinfo
		{author} {\bibfnamefont {E.}~\bibnamefont {Paris}}, \bibinfo {author}
		{\bibfnamefont {P.}~\bibnamefont {Puphal}}, \bibinfo {author} {\bibfnamefont
			{R.}~\bibnamefont {Bag}}, \bibinfo {author} {\bibfnamefont {G.}~\bibnamefont
			{Deng}}, \bibinfo {author} {\bibfnamefont {T.~C.}\ \bibnamefont {Asmara}},
		\bibinfo {author} {\bibfnamefont {V.~N.}\ \bibnamefont {Strocov}}, \bibinfo
		{author} {\bibfnamefont {S.}~\bibnamefont {Singh}}, \bibinfo {author}
		{\bibfnamefont {E.}~\bibnamefont {Pomjakushina}}, \bibinfo {author}
		{\bibfnamefont {U.}~\bibnamefont {Kumar}}, \bibinfo {author} {\bibfnamefont
			{A.}~\bibnamefont {Nocera}}, \bibinfo {author} {\bibfnamefont {H.~M.}\
			\bibnamefont {R{\o}nnow}}, \bibinfo {author} {\bibfnamefont {S.}~\bibnamefont
			{Johnston}},\ and\ \bibinfo {author} {\bibfnamefont {T.}~\bibnamefont
			{Schmitt}},\ }\bibfield  {title} {\bibinfo {title} {Crossover of high-energy
			spin fluctuations from collective triplons to localized magnetic excitations
			in {Sr}$_{14-x}${Ca}$_x${Cu}$_{24}${O}$_{41}$ ladders},\ }\href
	{https://doi.org/10.1038/s41535-022-00502-1} {\bibfield  {journal} {\bibinfo
			{journal} {npj Quantum Mater.}\ }\textbf {\bibinfo {volume} {7}},\ \bibinfo
		{pages} {92} (\bibinfo {year} {2022})}\BibitemShut {NoStop}%
	\bibitem [{\citenamefont {Scheie}\ \emph {et~al.}(2024)\citenamefont {Scheie},
		\citenamefont {Laurell}, \citenamefont {Thomas}, \citenamefont {Sharma},
		\citenamefont {Kolesnikov}, \citenamefont {Granroth}, \citenamefont {Zhang},
		\citenamefont {Lake}, \citenamefont {{Mihalik Jr.}}, \citenamefont {Bewley},
		\citenamefont {Eccleston}, \citenamefont {Akimitsu}, \citenamefont {Dagotto},
		\citenamefont {Batista}, \citenamefont {Alvarez}, \citenamefont {Johnston},\
		and\ \citenamefont {Tennant}}]{Scheie2024Cuprate}%
	\BibitemOpen
	\bibfield  {author} {\bibinfo {author} {\bibfnamefont {A.}~\bibnamefont
			{Scheie}}, \bibinfo {author} {\bibfnamefont {P.}~\bibnamefont {Laurell}},
		\bibinfo {author} {\bibfnamefont {J.}~\bibnamefont {Thomas}}, \bibinfo
		{author} {\bibfnamefont {V.}~\bibnamefont {Sharma}}, \bibinfo {author}
		{\bibfnamefont {A.~I.}\ \bibnamefont {Kolesnikov}}, \bibinfo {author}
		{\bibfnamefont {G.~E.}\ \bibnamefont {Granroth}}, \bibinfo {author}
		{\bibfnamefont {Q.}~\bibnamefont {Zhang}}, \bibinfo {author} {\bibfnamefont
			{B.}~\bibnamefont {Lake}}, \bibinfo {author} {\bibfnamefont {M.}~\bibnamefont
			{{Mihalik Jr.}}}, \bibinfo {author} {\bibfnamefont {R.~I.}\ \bibnamefont
			{Bewley}}, \bibinfo {author} {\bibfnamefont {R.~S.}\ \bibnamefont
			{Eccleston}}, \bibinfo {author} {\bibfnamefont {J.}~\bibnamefont {Akimitsu}},
		\bibinfo {author} {\bibfnamefont {E.}~\bibnamefont {Dagotto}}, \bibinfo
		{author} {\bibfnamefont {C.~D.}\ \bibnamefont {Batista}}, \bibinfo {author}
		{\bibfnamefont {G.}~\bibnamefont {Alvarez}}, \bibinfo {author} {\bibfnamefont
			{S.}~\bibnamefont {Johnston}},\ and\ \bibinfo {author} {\bibfnamefont
			{D.~A.}\ \bibnamefont {Tennant}},\ }\href@noop {} {\bibinfo {title}
		{Cooper-pair localization in the magnetic dynamics of a cuprate ladder}},\
	\bibinfo {howpublished} {(unpublished)} (\bibinfo {year} {2024})\BibitemShut
	{NoStop}%
	\bibitem [{\citenamefont {{Padma \it et al.}}(2024)}]{Padma2024Cuprate}%
	\BibitemOpen
	\bibfield  {author} {\bibinfo {author} {\bibfnamefont {H.}~\bibnamefont
			{{Padma \it et al.}}},\ }\href@noop {} {\bibinfo {title} {Beyond-{H}ubbard
			pairing in a cuprate ladder}},\ \bibinfo {howpublished} {(private
		communication)} (\bibinfo {year} {2024})\BibitemShut {NoStop}%
	\bibitem [{\citenamefont {Uehara}\ \emph {et~al.}(1996)\citenamefont {Uehara},
		\citenamefont {Nagata}, \citenamefont {Akimitsu}, \citenamefont {Takahashi},
		\citenamefont {Môri},\ and\ \citenamefont
		{Kinoshita}}]{doi:10.1143/JPSJ.65.2764}%
	\BibitemOpen
	\bibfield  {author} {\bibinfo {author} {\bibfnamefont {M.}~\bibnamefont
			{Uehara}}, \bibinfo {author} {\bibfnamefont {T.}~\bibnamefont {Nagata}},
		\bibinfo {author} {\bibfnamefont {J.}~\bibnamefont {Akimitsu}}, \bibinfo
		{author} {\bibfnamefont {H.}~\bibnamefont {Takahashi}}, \bibinfo {author}
		{\bibfnamefont {N.}~\bibnamefont {Môri}},\ and\ \bibinfo {author}
		{\bibfnamefont {K.}~\bibnamefont {Kinoshita}},\ }\bibfield  {title} {\bibinfo
		{title} {Superconductivity in the ladder material
			{Sr}$_{0.4}${Ca}$_{13.6}${Cu}$_{24}${O}$_{41.84}$},\ }\href
	{https://doi.org/10.1143/JPSJ.65.2764} {\bibfield  {journal} {\bibinfo
			{journal} {J. Phys. Soc. Jpn.}\ }\textbf {\bibinfo {volume} {65}},\ \bibinfo
		{pages} {2764} (\bibinfo {year} {1996})}\BibitemShut {NoStop}%
	\bibitem [{\citenamefont {Piskunov}\ \emph {et~al.}(2005)\citenamefont
		{Piskunov}, \citenamefont {J\'erome}, \citenamefont {Auban-Senzier},
		\citenamefont {Wzietek},\ and\ \citenamefont
		{Yakubovsky}}]{PhysRevB.72.064512}%
	\BibitemOpen
	\bibfield  {author} {\bibinfo {author} {\bibfnamefont {Y.}~\bibnamefont
			{Piskunov}}, \bibinfo {author} {\bibfnamefont {D.}~\bibnamefont {J\'erome}},
		\bibinfo {author} {\bibfnamefont {P.}~\bibnamefont {Auban-Senzier}}, \bibinfo
		{author} {\bibfnamefont {P.}~\bibnamefont {Wzietek}},\ and\ \bibinfo {author}
		{\bibfnamefont {A.}~\bibnamefont {Yakubovsky}},\ }\bibfield  {title}
	{\bibinfo {title} {Hole redistribution in
			{${\mathrm{Sr}}_{14\ensuremath{-}x}{\mathrm{Ca}}_{x}{\mathrm{Cu}}_{24}{\mathrm{O}}_{41}$
				$(x=0,12)$} spin ladder compounds: {$^{63}\mathrm{Cu}$ and $^{17}\mathrm{O}$
				NMR} studies under pressure},\ }\href
	{https://doi.org/10.1103/PhysRevB.72.064512} {\bibfield  {journal} {\bibinfo
			{journal} {Phys. Rev. B}\ }\textbf {\bibinfo {volume} {72}},\ \bibinfo
		{pages} {064512} (\bibinfo {year} {2005})}\BibitemShut {NoStop}%
	\bibitem [{\citenamefont {Takahashi}\ \emph {et~al.}(2015)\citenamefont
		{Takahashi}, \citenamefont {Sugimoto}, \citenamefont {Nambu}, \citenamefont
		{Yamauchi}, \citenamefont {Hirata}, \citenamefont {Kawakami}, \citenamefont
		{Avdeev}, \citenamefont {Matsubayashi}, \citenamefont {Du}, \citenamefont
		{Kawashima}, \citenamefont {Soeda}, \citenamefont {Nakano}, \citenamefont
		{Uwatoko}, \citenamefont {Ueda}, \citenamefont {Sato},\ and\ \citenamefont
		{Ohgushi}}]{Takahashi2015}%
	\BibitemOpen
	\bibfield  {author} {\bibinfo {author} {\bibfnamefont {H.}~\bibnamefont
			{Takahashi}}, \bibinfo {author} {\bibfnamefont {A.}~\bibnamefont {Sugimoto}},
		\bibinfo {author} {\bibfnamefont {Y.}~\bibnamefont {Nambu}}, \bibinfo
		{author} {\bibfnamefont {T.}~\bibnamefont {Yamauchi}}, \bibinfo {author}
		{\bibfnamefont {Y.}~\bibnamefont {Hirata}}, \bibinfo {author} {\bibfnamefont
			{T.}~\bibnamefont {Kawakami}}, \bibinfo {author} {\bibfnamefont
			{M.}~\bibnamefont {Avdeev}}, \bibinfo {author} {\bibfnamefont
			{K.}~\bibnamefont {Matsubayashi}}, \bibinfo {author} {\bibfnamefont
			{F.}~\bibnamefont {Du}}, \bibinfo {author} {\bibfnamefont {C.}~\bibnamefont
			{Kawashima}}, \bibinfo {author} {\bibfnamefont {H.}~\bibnamefont {Soeda}},
		\bibinfo {author} {\bibfnamefont {S.}~\bibnamefont {Nakano}}, \bibinfo
		{author} {\bibfnamefont {Y.}~\bibnamefont {Uwatoko}}, \bibinfo {author}
		{\bibfnamefont {Y.}~\bibnamefont {Ueda}}, \bibinfo {author} {\bibfnamefont
			{T.~J.}\ \bibnamefont {Sato}},\ and\ \bibinfo {author} {\bibfnamefont
			{K.}~\bibnamefont {Ohgushi}},\ }\bibfield  {title} {\bibinfo {title}
		{Pressure-induced superconductivity in the iron-based ladder material
			{BaFe}$_2${S}$_3$},\ }\href
	{https://doi.org/https://doi.org/10.1038/nmat4351} {\bibfield  {journal}
		{\bibinfo  {journal} {Nat. Mater.}\ }\textbf {\bibinfo {volume} {14}},\
		\bibinfo {pages} {1008} (\bibinfo {year} {2015})}\BibitemShut {NoStop}%
	\bibitem [{\citenamefont {Yamauchi}\ \emph {et~al.}(2015)\citenamefont
		{Yamauchi}, \citenamefont {Hirata}, \citenamefont {Ueda},\ and\ \citenamefont
		{Ohgushi}}]{Yamauchi2015}%
	\BibitemOpen
	\bibfield  {author} {\bibinfo {author} {\bibfnamefont {T.}~\bibnamefont
			{Yamauchi}}, \bibinfo {author} {\bibfnamefont {Y.}~\bibnamefont {Hirata}},
		\bibinfo {author} {\bibfnamefont {Y.}~\bibnamefont {Ueda}},\ and\ \bibinfo
		{author} {\bibfnamefont {K.}~\bibnamefont {Ohgushi}},\ }\bibfield  {title}
	{\bibinfo {title} {Pressure-induced {Mott} transition followed by a 24-{K}
			superconducting phase in {${\mathrm{BaFe}}_{2}{\mathrm{S}}_{3}$}},\ }\href
	{https://doi.org/10.1103/PhysRevLett.115.246402} {\bibfield  {journal}
		{\bibinfo  {journal} {Phys. Rev. Lett.}\ }\textbf {\bibinfo {volume} {115}},\
		\bibinfo {pages} {246402} (\bibinfo {year} {2015})}\BibitemShut {NoStop}%
	\bibitem [{\citenamefont {Sous}\ \emph {et~al.}(2021)\citenamefont {Sous},
		\citenamefont {Gadjieva}, \citenamefont {Nuckolls}, \citenamefont
		{Reichman},\ and\ \citenamefont {Millis}}]{doi:10.1021/acs.nanolett.1c03236}%
	\BibitemOpen
	\bibfield  {author} {\bibinfo {author} {\bibfnamefont {J.}~\bibnamefont
			{Sous}}, \bibinfo {author} {\bibfnamefont {N.~A.}\ \bibnamefont {Gadjieva}},
		\bibinfo {author} {\bibfnamefont {C.}~\bibnamefont {Nuckolls}}, \bibinfo
		{author} {\bibfnamefont {D.~R.}\ \bibnamefont {Reichman}},\ and\ \bibinfo
		{author} {\bibfnamefont {A.~J.}\ \bibnamefont {Millis}},\ }\bibfield  {title}
	{\bibinfo {title} {Strongly correlated ladders in {K}-doped p-{Terphenyl}
			crystals},\ }\href {https://doi.org/10.1021/acs.nanolett.1c03236} {\bibfield
		{journal} {\bibinfo  {journal} {Nano Lett.}\ }\textbf {\bibinfo {volume}
			{21}},\ \bibinfo {pages} {9573} (\bibinfo {year} {2021})}\BibitemShut
	{NoStop}%
	\bibitem [{\citenamefont {Hirthe}\ \emph {et~al.}(2023)\citenamefont {Hirthe},
		\citenamefont {Chalopin}, \citenamefont {Bourgund}, \citenamefont {Bojović},
		\citenamefont {Bohrdt}, \citenamefont {Demler}, \citenamefont {Grusdt},
		\citenamefont {Bloch},\ and\ \citenamefont {Hilker}}]{Hirthe2023}%
	\BibitemOpen
	\bibfield  {author} {\bibinfo {author} {\bibfnamefont {S.}~\bibnamefont
			{Hirthe}}, \bibinfo {author} {\bibfnamefont {T.}~\bibnamefont {Chalopin}},
		\bibinfo {author} {\bibfnamefont {D.}~\bibnamefont {Bourgund}}, \bibinfo
		{author} {\bibfnamefont {P.}~\bibnamefont {Bojović}}, \bibinfo {author}
		{\bibfnamefont {A.}~\bibnamefont {Bohrdt}}, \bibinfo {author} {\bibfnamefont
			{E.}~\bibnamefont {Demler}}, \bibinfo {author} {\bibfnamefont
			{F.}~\bibnamefont {Grusdt}}, \bibinfo {author} {\bibfnamefont
			{I.}~\bibnamefont {Bloch}},\ and\ \bibinfo {author} {\bibfnamefont {T.~A.}\
			\bibnamefont {Hilker}},\ }\bibfield  {title} {\bibinfo {title} {Magnetically
			mediated hole pairing in fermionic ladders of ultracold atoms},\ }\href
	{https://doi.org/10.1038/s41586-022-05437-y} {\bibfield  {journal} {\bibinfo
			{journal} {Nature}\ }\textbf {\bibinfo {volume} {613}},\ \bibinfo {pages}
		{463} (\bibinfo {year} {2023})}\BibitemShut {NoStop}%
	\bibitem [{\citenamefont {Xu}\ \emph {et~al.}(2000)\citenamefont {Xu},
		\citenamefont {Aeppli}, \citenamefont {Bisher}, \citenamefont {Broholm},
		\citenamefont {DiTusa}, \citenamefont {Frost}, \citenamefont {Ito},
		\citenamefont {Oka}, \citenamefont {Paul}, \citenamefont {Takagi},\ and\
		\citenamefont {Treacy}}]{doi:10.1126/science.289.5478.419}%
	\BibitemOpen
	\bibfield  {author} {\bibinfo {author} {\bibfnamefont {G.}~\bibnamefont
			{Xu}}, \bibinfo {author} {\bibfnamefont {G.}~\bibnamefont {Aeppli}}, \bibinfo
		{author} {\bibfnamefont {M.~E.}\ \bibnamefont {Bisher}}, \bibinfo {author}
		{\bibfnamefont {C.}~\bibnamefont {Broholm}}, \bibinfo {author} {\bibfnamefont
			{J.~F.}\ \bibnamefont {DiTusa}}, \bibinfo {author} {\bibfnamefont {C.~D.}\
			\bibnamefont {Frost}}, \bibinfo {author} {\bibfnamefont {T.}~\bibnamefont
			{Ito}}, \bibinfo {author} {\bibfnamefont {K.}~\bibnamefont {Oka}}, \bibinfo
		{author} {\bibfnamefont {R.~L.}\ \bibnamefont {Paul}}, \bibinfo {author}
		{\bibfnamefont {H.}~\bibnamefont {Takagi}},\ and\ \bibinfo {author}
		{\bibfnamefont {M.~M.~J.}\ \bibnamefont {Treacy}},\ }\bibfield  {title}
	{\bibinfo {title} {Holes in a quantum spin liquid},\ }\href
	{https://doi.org/10.1126/science.289.5478.419} {\bibfield  {journal}
		{\bibinfo  {journal} {Science}\ }\textbf {\bibinfo {volume} {289}},\ \bibinfo
		{pages} {419} (\bibinfo {year} {2000})}\BibitemShut {NoStop}%
	\bibitem [{\citenamefont {Ning}\ \emph {et~al.}(2020)\citenamefont {Ning},
		\citenamefont {Liu},\ and\ \citenamefont {Jiang}}]{PhysRevResearch.2.023184}%
	\BibitemOpen
	\bibfield  {author} {\bibinfo {author} {\bibfnamefont {S.-Q.}\ \bibnamefont
			{Ning}}, \bibinfo {author} {\bibfnamefont {Z.-X.}\ \bibnamefont {Liu}},\ and\
		\bibinfo {author} {\bibfnamefont {H.-C.}\ \bibnamefont {Jiang}},\ }\bibfield
	{title} {\bibinfo {title} {Topological superconductivity by doping
			symmetry-protected topological states},\ }\href
	{https://doi.org/10.1103/PhysRevResearch.2.023184} {\bibfield  {journal}
		{\bibinfo  {journal} {Phys. Rev. Research}\ }\textbf {\bibinfo {volume}
			{2}},\ \bibinfo {pages} {023184} (\bibinfo {year} {2020})}\BibitemShut
	{NoStop}%
	\bibitem [{\citenamefont {Zhang}\ and\ \citenamefont
		{Vishwanath}(2022)}]{PhysRevB.106.045103}%
	\BibitemOpen
	\bibfield  {author} {\bibinfo {author} {\bibfnamefont {Y.-H.}\ \bibnamefont
			{Zhang}}\ and\ \bibinfo {author} {\bibfnamefont {A.}~\bibnamefont
			{Vishwanath}},\ }\bibfield  {title} {\bibinfo {title} {Pair-density-wave
			superconductor from doping {Haldane} chain and rung-singlet ladder},\ }\href
	{https://doi.org/10.1103/PhysRevB.106.045103} {\bibfield  {journal} {\bibinfo
			{journal} {Phys. Rev. B}\ }\textbf {\bibinfo {volume} {106}},\ \bibinfo
		{pages} {045103} (\bibinfo {year} {2022})}\BibitemShut {NoStop}%
	\bibitem [{\citenamefont {Haldane}(1983)}]{PhysRevLett.50.1153}%
	\BibitemOpen
	\bibfield  {author} {\bibinfo {author} {\bibfnamefont {F.~D.~M.}\
			\bibnamefont {Haldane}},\ }\bibfield  {title} {\bibinfo {title} {Nonlinear
			field theory of large-spin {Heisenberg} antiferromagnets: Semiclassically
			quantized solitons of the one-dimensional easy-axis {N\'eel} state},\ }\href
	{https://doi.org/10.1103/PhysRevLett.50.1153} {\bibfield  {journal} {\bibinfo
			{journal} {Phys. Rev. Lett.}\ }\textbf {\bibinfo {volume} {50}},\ \bibinfo
		{pages} {1153} (\bibinfo {year} {1983})}\BibitemShut {NoStop}%
	\bibitem [{\citenamefont {Affleck}\ \emph {et~al.}(1987)\citenamefont
		{Affleck}, \citenamefont {Kennedy}, \citenamefont {Lieb},\ and\ \citenamefont
		{Tasaki}}]{PhysRevLett.59.799}%
	\BibitemOpen
	\bibfield  {author} {\bibinfo {author} {\bibfnamefont {I.}~\bibnamefont
			{Affleck}}, \bibinfo {author} {\bibfnamefont {T.}~\bibnamefont {Kennedy}},
		\bibinfo {author} {\bibfnamefont {E.~H.}\ \bibnamefont {Lieb}},\ and\
		\bibinfo {author} {\bibfnamefont {H.}~\bibnamefont {Tasaki}},\ }\bibfield
	{title} {\bibinfo {title} {Rigorous results on valence-bond ground states in
			antiferromagnets},\ }\href {https://doi.org/10.1103/PhysRevLett.59.799}
	{\bibfield  {journal} {\bibinfo  {journal} {Phys. Rev. Lett.}\ }\textbf
		{\bibinfo {volume} {59}},\ \bibinfo {pages} {799} (\bibinfo {year}
		{1987})}\BibitemShut {NoStop}%
	\bibitem [{\citenamefont {Affleck}(1989)}]{Affleck1989a}%
	\BibitemOpen
	\bibfield  {author} {\bibinfo {author} {\bibfnamefont {I.}~\bibnamefont
			{Affleck}},\ }\bibfield  {title} {\bibinfo {title} {Quantum spin chains and
			the {Haldane} gap},\ }\href {https://doi.org/10.1088/0953-8984/1/19/001}
	{\bibfield  {journal} {\bibinfo  {journal} {J. Phys. Condens. Matter}\
		}\textbf {\bibinfo {volume} {1}},\ \bibinfo {pages} {3047} (\bibinfo {year}
		{1989})}\BibitemShut {NoStop}%
	\bibitem [{\citenamefont {Gu}\ and\ \citenamefont
		{Wen}(2009)}]{PhysRevB.80.155131}%
	\BibitemOpen
	\bibfield  {author} {\bibinfo {author} {\bibfnamefont {Z.-C.}\ \bibnamefont
			{Gu}}\ and\ \bibinfo {author} {\bibfnamefont {X.-G.}\ \bibnamefont {Wen}},\
	}\bibfield  {title} {\bibinfo {title} {Tensor-entanglement-filtering
			renormalization approach and symmetry-protected topological order},\ }\href
	{https://doi.org/10.1103/PhysRevB.80.155131} {\bibfield  {journal} {\bibinfo
			{journal} {Phys. Rev. B}\ }\textbf {\bibinfo {volume} {80}},\ \bibinfo
		{pages} {155131} (\bibinfo {year} {2009})}\BibitemShut {NoStop}%
	\bibitem [{\citenamefont {Pollmann}\ \emph {et~al.}(2010)\citenamefont
		{Pollmann}, \citenamefont {Turner}, \citenamefont {Berg},\ and\ \citenamefont
		{Oshikawa}}]{PhysRevB.81.064439}%
	\BibitemOpen
	\bibfield  {author} {\bibinfo {author} {\bibfnamefont {F.}~\bibnamefont
			{Pollmann}}, \bibinfo {author} {\bibfnamefont {A.~M.}\ \bibnamefont
			{Turner}}, \bibinfo {author} {\bibfnamefont {E.}~\bibnamefont {Berg}},\ and\
		\bibinfo {author} {\bibfnamefont {M.}~\bibnamefont {Oshikawa}},\ }\bibfield
	{title} {\bibinfo {title} {Entanglement spectrum of a topological phase in
			one dimension},\ }\href {https://doi.org/10.1103/PhysRevB.81.064439}
	{\bibfield  {journal} {\bibinfo  {journal} {Phys. Rev. B}\ }\textbf {\bibinfo
			{volume} {81}},\ \bibinfo {pages} {064439} (\bibinfo {year}
		{2010})}\BibitemShut {NoStop}%
	\bibitem [{\citenamefont {Pollmann}\ \emph {et~al.}(2012)\citenamefont
		{Pollmann}, \citenamefont {Berg}, \citenamefont {Turner},\ and\ \citenamefont
		{Oshikawa}}]{PhysRevB.85.075125}%
	\BibitemOpen
	\bibfield  {author} {\bibinfo {author} {\bibfnamefont {F.}~\bibnamefont
			{Pollmann}}, \bibinfo {author} {\bibfnamefont {E.}~\bibnamefont {Berg}},
		\bibinfo {author} {\bibfnamefont {A.~M.}\ \bibnamefont {Turner}},\ and\
		\bibinfo {author} {\bibfnamefont {M.}~\bibnamefont {Oshikawa}},\ }\bibfield
	{title} {\bibinfo {title} {Symmetry protection of topological phases in
			one-dimensional quantum spin systems},\ }\href
	{https://doi.org/10.1103/PhysRevB.85.075125} {\bibfield  {journal} {\bibinfo
			{journal} {Phys. Rev. B}\ }\textbf {\bibinfo {volume} {85}},\ \bibinfo
		{pages} {075125} (\bibinfo {year} {2012})}\BibitemShut {NoStop}%
	\bibitem [{\citenamefont {den Nijs}\ and\ \citenamefont
		{Rommelse}(1989)}]{PhysRevB.40.4709}%
	\BibitemOpen
	\bibfield  {author} {\bibinfo {author} {\bibfnamefont {M.}~\bibnamefont {den
				Nijs}}\ and\ \bibinfo {author} {\bibfnamefont {K.}~\bibnamefont {Rommelse}},\
	}\bibfield  {title} {\bibinfo {title} {Preroughening transitions in crystal
			surfaces and valence-bond phases in quantum spin chains},\ }\href
	{https://doi.org/10.1103/PhysRevB.40.4709} {\bibfield  {journal} {\bibinfo
			{journal} {Phys. Rev. B}\ }\textbf {\bibinfo {volume} {40}},\ \bibinfo
		{pages} {4709} (\bibinfo {year} {1989})}\BibitemShut {NoStop}%
	\bibitem [{\citenamefont {Kennedy}\ and\ \citenamefont
		{Tasaki}(1992)}]{Kennedy1992}%
	\BibitemOpen
	\bibfield  {author} {\bibinfo {author} {\bibfnamefont {T.}~\bibnamefont
			{Kennedy}}\ and\ \bibinfo {author} {\bibfnamefont {H.}~\bibnamefont
			{Tasaki}},\ }\bibfield  {title} {\bibinfo {title} {Hidden symmetry breaking
			and the {Haldane} phase in {$S=1$} quantum spin chains},\ }\href
	{https://doi.org/https://doi.org/10.1007/BF02097239} {\bibfield  {journal}
		{\bibinfo  {journal} {Commun. Math. Phys.}\ }\textbf {\bibinfo {volume}
			{147}},\ \bibinfo {pages} {431} (\bibinfo {year} {1992})}\BibitemShut
	{NoStop}%
	\bibitem [{\citenamefont {Riera}\ \emph {et~al.}(1997)\citenamefont {Riera},
		\citenamefont {Hallberg},\ and\ \citenamefont
		{Dagotto}}]{PhysRevLett.79.713}%
	\BibitemOpen
	\bibfield  {author} {\bibinfo {author} {\bibfnamefont {J.}~\bibnamefont
			{Riera}}, \bibinfo {author} {\bibfnamefont {K.}~\bibnamefont {Hallberg}},\
		and\ \bibinfo {author} {\bibfnamefont {E.}~\bibnamefont {Dagotto}},\
	}\bibfield  {title} {\bibinfo {title} {Phase diagram of electronic models for
			transition metal oxides in one dimension},\ }\href
	{https://doi.org/10.1103/PhysRevLett.79.713} {\bibfield  {journal} {\bibinfo
			{journal} {Phys. Rev. Lett.}\ }\textbf {\bibinfo {volume} {79}},\ \bibinfo
		{pages} {713} (\bibinfo {year} {1997})}\BibitemShut {NoStop}%
	\bibitem [{\citenamefont {Ammon}\ and\ \citenamefont
		{Imada}(2000)}]{doi:10.1143/JPSJ.69.1946}%
	\BibitemOpen
	\bibfield  {author} {\bibinfo {author} {\bibfnamefont {B.}~\bibnamefont
			{Ammon}}\ and\ \bibinfo {author} {\bibfnamefont {M.}~\bibnamefont {Imada}},\
	}\bibfield  {title} {\bibinfo {title} {Effect of the orbital level difference
			in doped spin-1 chains},\ }\href {https://doi.org/10.1143/JPSJ.69.1946}
	{\bibfield  {journal} {\bibinfo  {journal} {J. Phys. Soc. Jpn.}\ }\textbf
		{\bibinfo {volume} {69}},\ \bibinfo {pages} {1946} (\bibinfo {year}
		{2000})}\BibitemShut {NoStop}%
	\bibitem [{\citenamefont {Ammon}\ and\ \citenamefont
		{Imada}(2001)}]{doi:10.1143/JPSJ.70.547}%
	\BibitemOpen
	\bibfield  {author} {\bibinfo {author} {\bibfnamefont {B.}~\bibnamefont
			{Ammon}}\ and\ \bibinfo {author} {\bibfnamefont {M.}~\bibnamefont {Imada}},\
	}\bibfield  {title} {\bibinfo {title} {Doped two orbital chains with strong
			{Hund's} rule couplings - ferromagnetism, spin gap, singlet and triplet
			pairings},\ }\href {https://doi.org/10.1143/JPSJ.70.547} {\bibfield
		{journal} {\bibinfo  {journal} {J. Phys. Soc. Jpn.}\ }\textbf {\bibinfo
			{volume} {70}},\ \bibinfo {pages} {547} (\bibinfo {year} {2001})}\BibitemShut
	{NoStop}%
	\bibitem [{\citenamefont {Jiang}\ \emph {et~al.}(2018)\citenamefont {Jiang},
		\citenamefont {Li}, \citenamefont {Seidel},\ and\ \citenamefont
		{Lee}}]{Jiang2018}%
	\BibitemOpen
	\bibfield  {author} {\bibinfo {author} {\bibfnamefont {H.-C.}\ \bibnamefont
			{Jiang}}, \bibinfo {author} {\bibfnamefont {Z.-X.}\ \bibnamefont {Li}},
		\bibinfo {author} {\bibfnamefont {A.}~\bibnamefont {Seidel}},\ and\ \bibinfo
		{author} {\bibfnamefont {D.-H.}\ \bibnamefont {Lee}},\ }\bibfield  {title}
	{\bibinfo {title} {Symmetry protected topological {Luttinger} liquids and the
			phase transition between them},\ }\href
	{https://doi.org/https://doi.org/10.1016/j.scib.2018.05.010} {\bibfield
		{journal} {\bibinfo  {journal} {Sci. Bull.}\ }\textbf {\bibinfo {volume}
			{63}},\ \bibinfo {pages} {753} (\bibinfo {year} {2018})}\BibitemShut
	{NoStop}%
	\bibitem [{\citenamefont {Hou}\ \emph {et~al.}(2019)\citenamefont {Hou},
		\citenamefont {Lee}, \citenamefont {Lou},\ and\ \citenamefont
		{Chen}}]{PhysRevB.99.094510}%
	\BibitemOpen
	\bibfield  {author} {\bibinfo {author} {\bibfnamefont {J.}~\bibnamefont
			{Hou}}, \bibinfo {author} {\bibfnamefont {T.-K.}\ \bibnamefont {Lee}},
		\bibinfo {author} {\bibfnamefont {J.}~\bibnamefont {Lou}},\ and\ \bibinfo
		{author} {\bibfnamefont {Y.}~\bibnamefont {Chen}},\ }\bibfield  {title}
	{\bibinfo {title} {Emergence of ${d}_{xy}$-wave superconductivity in a doped
			two-leg diagonal ladder},\ }\href
	{https://doi.org/10.1103/PhysRevB.99.094510} {\bibfield  {journal} {\bibinfo
			{journal} {Phys. Rev. B}\ }\textbf {\bibinfo {volume} {99}},\ \bibinfo
		{pages} {094510} (\bibinfo {year} {2019})}\BibitemShut {NoStop}%
	\bibitem [{\citenamefont {Shirakawa}\ \emph {et~al.}(2007)\citenamefont
		{Shirakawa}, \citenamefont {Ohta},\ and\ \citenamefont
		{Nishimoto}}]{SHIRAKAWA2007663}%
	\BibitemOpen
	\bibfield  {author} {\bibinfo {author} {\bibfnamefont {T.}~\bibnamefont
			{Shirakawa}}, \bibinfo {author} {\bibfnamefont {Y.}~\bibnamefont {Ohta}},\
		and\ \bibinfo {author} {\bibfnamefont {S.}~\bibnamefont {Nishimoto}},\
	}\bibfield  {title} {\bibinfo {title} {Spin-triplet superconductivity in the
			double-chain {Hubbard} model with ferromagnetic exchange interaction},\
	}\href {https://doi.org/https://doi.org/10.1016/j.jmmm.2006.10.188}
	{\bibfield  {journal} {\bibinfo  {journal} {J. Magn. Magn. Mater.}\ }\textbf
		{\bibinfo {volume} {310}},\ \bibinfo {pages} {663} (\bibinfo {year}
		{2007})},\ \bibinfo {note} {{Proceedings} of the 17th International
		Conference on Magnetism}\BibitemShut {NoStop}%
	\bibitem [{\citenamefont {Nishimoto}\ \emph {et~al.}(2007)\citenamefont
		{Nishimoto}, \citenamefont {Shirakawa},\ and\ \citenamefont
		{Ohta}}]{NISHIMOTO20071059}%
	\BibitemOpen
	\bibfield  {author} {\bibinfo {author} {\bibfnamefont {S.}~\bibnamefont
			{Nishimoto}}, \bibinfo {author} {\bibfnamefont {T.}~\bibnamefont
			{Shirakawa}},\ and\ \bibinfo {author} {\bibfnamefont {Y.}~\bibnamefont
			{Ohta}},\ }\bibfield  {title} {\bibinfo {title} {Spin-triplet
			superconductivity in the {Hubbard} chains coupled with ferromagnetic exchange
			interaction},\ }\href
	{https://doi.org/https://doi.org/10.1016/j.physc.2007.03.216} {\bibfield
		{journal} {\bibinfo  {journal} {Physica C: Supercond.}\ }\textbf {\bibinfo
			{volume} {460-462}},\ \bibinfo {pages} {1059} (\bibinfo {year} {2007})},\
	\bibinfo {note} {{P}roceedings of the 8th International Conference on
		Materials and Mechanisms of Superconductivity and High Temperature
		Superconductors}\BibitemShut {NoStop}%
	\bibitem [{\citenamefont {Shirakawa}\ \emph {et~al.}(2008)\citenamefont
		{Shirakawa}, \citenamefont {Nishimoto},\ and\ \citenamefont
		{Ohta}}]{PhysRevB.77.224510}%
	\BibitemOpen
	\bibfield  {author} {\bibinfo {author} {\bibfnamefont {T.}~\bibnamefont
			{Shirakawa}}, \bibinfo {author} {\bibfnamefont {S.}~\bibnamefont
			{Nishimoto}},\ and\ \bibinfo {author} {\bibfnamefont {Y.}~\bibnamefont
			{Ohta}},\ }\bibfield  {title} {\bibinfo {title} {Superconductivity in a model
			of two {Hubbard} chains coupled with ferromagnetic exchange interaction},\
	}\href {https://doi.org/10.1103/PhysRevB.77.224510} {\bibfield  {journal}
		{\bibinfo  {journal} {Phys. Rev. B}\ }\textbf {\bibinfo {volume} {77}},\
		\bibinfo {pages} {224510} (\bibinfo {year} {2008})}\BibitemShut {NoStop}%
	\bibitem [{\citenamefont {Fujimoto}\ and\ \citenamefont
		{Kawakami}(1995)}]{PhysRevB.52.6189}%
	\BibitemOpen
	\bibfield  {author} {\bibinfo {author} {\bibfnamefont {S.}~\bibnamefont
			{Fujimoto}}\ and\ \bibinfo {author} {\bibfnamefont {N.}~\bibnamefont
			{Kawakami}},\ }\bibfield  {title} {\bibinfo {title} {Weak-coupling approach
			to hole-doped {S}=1 {Haldane} systems},\ }\href
	{https://doi.org/10.1103/PhysRevB.52.6189} {\bibfield  {journal} {\bibinfo
			{journal} {Phys. Rev. B}\ }\textbf {\bibinfo {volume} {52}},\ \bibinfo
		{pages} {6189} (\bibinfo {year} {1995})}\BibitemShut {NoStop}%
	\bibitem [{\citenamefont {Shelton}\ and\ \citenamefont
		{Tsvelik}(1996)}]{PhysRevB.53.14036}%
	\BibitemOpen
	\bibfield  {author} {\bibinfo {author} {\bibfnamefont {D.~G.}\ \bibnamefont
			{Shelton}}\ and\ \bibinfo {author} {\bibfnamefont {A.~M.}\ \bibnamefont
			{Tsvelik}},\ }\bibfield  {title} {\bibinfo {title} {Superconductivity in a
			spin liquid: A one-dimensional example},\ }\href
	{https://doi.org/10.1103/PhysRevB.53.14036} {\bibfield  {journal} {\bibinfo
			{journal} {Phys. Rev. B}\ }\textbf {\bibinfo {volume} {53}},\ \bibinfo
		{pages} {14036} (\bibinfo {year} {1996})}\BibitemShut {NoStop}%
	\bibitem [{\citenamefont {Nagaosa}\ and\ \citenamefont
		{Oshikawa}(1996)}]{doi:10.1143/JPSJ.65.2241}%
	\BibitemOpen
	\bibfield  {author} {\bibinfo {author} {\bibfnamefont {N.}~\bibnamefont
			{Nagaosa}}\ and\ \bibinfo {author} {\bibfnamefont {M.}~\bibnamefont
			{Oshikawa}},\ }\bibfield  {title} {\bibinfo {title} {Chiral anomaly and spin
			gap in one-dimensional interacting fermions},\ }\href
	{https://doi.org/10.1143/JPSJ.65.2241} {\bibfield  {journal} {\bibinfo
			{journal} {J. Phys. Soc. Jpn.}\ }\textbf {\bibinfo {volume} {65}},\ \bibinfo
		{pages} {2241} (\bibinfo {year} {1996})}\BibitemShut {NoStop}%
	\bibitem [{\citenamefont {Patel}\ \emph {et~al.}(2020)\citenamefont {Patel},
		\citenamefont {Kaushal}, \citenamefont {Nocera}, \citenamefont {Alvarez},\
		and\ \citenamefont {Dagotto}}]{Patel2020}%
	\BibitemOpen
	\bibfield  {author} {\bibinfo {author} {\bibfnamefont {N.~D.}\ \bibnamefont
			{Patel}}, \bibinfo {author} {\bibfnamefont {N.}~\bibnamefont {Kaushal}},
		\bibinfo {author} {\bibfnamefont {A.}~\bibnamefont {Nocera}}, \bibinfo
		{author} {\bibfnamefont {G.}~\bibnamefont {Alvarez}},\ and\ \bibinfo {author}
		{\bibfnamefont {E.}~\bibnamefont {Dagotto}},\ }\bibfield  {title} {\bibinfo
		{title} {Emergence of superconductivity in doped multiorbital {Hubbard}
			chains},\ }\href {https://doi.org/https://doi.org/10.1038/s41535-020-0228-2}
	{\bibfield  {journal} {\bibinfo  {journal} {npj Quantum Mater.}\ }\textbf
		{\bibinfo {volume} {5}},\ \bibinfo {pages} {27} (\bibinfo {year}
		{2020})}\BibitemShut {NoStop}%
	\bibitem [{\citenamefont {Patel}\ \emph {et~al.}(2017)\citenamefont {Patel},
		\citenamefont {Nocera}, \citenamefont {Alvarez}, \citenamefont {Moreo},\ and\
		\citenamefont {Dagotto}}]{PhysRevB.96.024520}%
	\BibitemOpen
	\bibfield  {author} {\bibinfo {author} {\bibfnamefont {N.~D.}\ \bibnamefont
			{Patel}}, \bibinfo {author} {\bibfnamefont {A.}~\bibnamefont {Nocera}},
		\bibinfo {author} {\bibfnamefont {G.}~\bibnamefont {Alvarez}}, \bibinfo
		{author} {\bibfnamefont {A.}~\bibnamefont {Moreo}},\ and\ \bibinfo {author}
		{\bibfnamefont {E.}~\bibnamefont {Dagotto}},\ }\bibfield  {title} {\bibinfo
		{title} {Pairing tendencies in a two-orbital {Hubbard} model in one
			dimension},\ }\href {https://doi.org/10.1103/PhysRevB.96.024520} {\bibfield
		{journal} {\bibinfo  {journal} {Phys. Rev. B}\ }\textbf {\bibinfo {volume}
			{96}},\ \bibinfo {pages} {024520} (\bibinfo {year} {2017})}\BibitemShut
	{NoStop}%
	\bibitem [{\citenamefont {Li}\ and\ \citenamefont
		{Haldane}(2008)}]{PhysRevLett.101.010504}%
	\BibitemOpen
	\bibfield  {author} {\bibinfo {author} {\bibfnamefont {H.}~\bibnamefont
			{Li}}\ and\ \bibinfo {author} {\bibfnamefont {F.~D.~M.}\ \bibnamefont
			{Haldane}},\ }\bibfield  {title} {\bibinfo {title} {Entanglement spectrum as
			a generalization of entanglement entropy: Identification of topological order
			in non-{Abelian} fractional quantum {Hall} effect states},\ }\href
	{https://doi.org/10.1103/PhysRevLett.101.010504} {\bibfield  {journal}
		{\bibinfo  {journal} {Phys. Rev. Lett.}\ }\textbf {\bibinfo {volume} {101}},\
		\bibinfo {pages} {010504} (\bibinfo {year} {2008})}\BibitemShut {NoStop}%
	\bibitem [{\citenamefont {Jażdżewska}\ \emph {et~al.}(2023)\citenamefont
		{Jażdżewska}, \citenamefont {Mierzejewski}, \citenamefont {Środa},
		\citenamefont {Nocera}, \citenamefont {Alvarez}, \citenamefont {Dagotto},\
		and\ \citenamefont {Herbrych}}]{Jazdzewska2023}%
	\BibitemOpen
	\bibfield  {author} {\bibinfo {author} {\bibfnamefont {A.}~\bibnamefont
			{Jażdżewska}}, \bibinfo {author} {\bibfnamefont {M.}~\bibnamefont
			{Mierzejewski}}, \bibinfo {author} {\bibfnamefont {M.}~\bibnamefont
			{Środa}}, \bibinfo {author} {\bibfnamefont {A.}~\bibnamefont {Nocera}},
		\bibinfo {author} {\bibfnamefont {G.}~\bibnamefont {Alvarez}}, \bibinfo
		{author} {\bibfnamefont {E.}~\bibnamefont {Dagotto}},\ and\ \bibinfo {author}
		{\bibfnamefont {J.}~\bibnamefont {Herbrych}},\ }\bibfield  {title} {\bibinfo
		{title} {Transition to the {Haldane} phase driven by electron-electron
			correlations},\ }\href
	{https://doi.org/https://doi.org/10.1038/s41467-023-44135-9} {\bibfield
		{journal} {\bibinfo  {journal} {Nat. Commun.}\ }\textbf {\bibinfo {volume}
			{14}},\ \bibinfo {pages} {8524} (\bibinfo {year} {2023})}\BibitemShut
	{NoStop}%
	\bibitem [{\citenamefont {Luther}\ and\ \citenamefont
		{Emery}(1974)}]{PhysRevLett.33.589}%
	\BibitemOpen
	\bibfield  {author} {\bibinfo {author} {\bibfnamefont {A.}~\bibnamefont
			{Luther}}\ and\ \bibinfo {author} {\bibfnamefont {V.~J.}\ \bibnamefont
			{Emery}},\ }\bibfield  {title} {\bibinfo {title} {Backward scattering in the
			one-dimensional electron gas},\ }\href
	{https://doi.org/10.1103/PhysRevLett.33.589} {\bibfield  {journal} {\bibinfo
			{journal} {Phys. Rev. Lett.}\ }\textbf {\bibinfo {volume} {33}},\ \bibinfo
		{pages} {589} (\bibinfo {year} {1974})}\BibitemShut {NoStop}%
	\bibitem [{\citenamefont {Luo}\ \emph {et~al.}(2010)\citenamefont {Luo},
		\citenamefont {Martins}, \citenamefont {Yao}, \citenamefont {Daghofer},
		\citenamefont {Yu}, \citenamefont {Moreo},\ and\ \citenamefont
		{Dagotto}}]{PhysRevB.82.104508}%
	\BibitemOpen
	\bibfield  {author} {\bibinfo {author} {\bibfnamefont {Q.}~\bibnamefont
			{Luo}}, \bibinfo {author} {\bibfnamefont {G.}~\bibnamefont {Martins}},
		\bibinfo {author} {\bibfnamefont {D.-X.}\ \bibnamefont {Yao}}, \bibinfo
		{author} {\bibfnamefont {M.}~\bibnamefont {Daghofer}}, \bibinfo {author}
		{\bibfnamefont {R.}~\bibnamefont {Yu}}, \bibinfo {author} {\bibfnamefont
			{A.}~\bibnamefont {Moreo}},\ and\ \bibinfo {author} {\bibfnamefont
			{E.}~\bibnamefont {Dagotto}},\ }\bibfield  {title} {\bibinfo {title} {Neutron
			and {ARPES} constraints on the couplings of the multiorbital {Hubbard} model
			for the iron pnictides},\ }\href {https://doi.org/10.1103/PhysRevB.82.104508}
	{\bibfield  {journal} {\bibinfo  {journal} {Phys. Rev. B}\ }\textbf {\bibinfo
			{volume} {82}},\ \bibinfo {pages} {104508} (\bibinfo {year}
		{2010})}\BibitemShut {NoStop}%
	\bibitem [{\citenamefont {Dai}\ \emph {et~al.}(2012)\citenamefont {Dai},
		\citenamefont {Hu},\ and\ \citenamefont {Dagotto}}]{Dai2012}%
	\BibitemOpen
	\bibfield  {author} {\bibinfo {author} {\bibfnamefont {P.}~\bibnamefont
			{Dai}}, \bibinfo {author} {\bibfnamefont {J.}~\bibnamefont {Hu}},\ and\
		\bibinfo {author} {\bibfnamefont {E.}~\bibnamefont {Dagotto}},\ }\bibfield
	{title} {\bibinfo {title} {Magnetism and its microscopic origin in iron-based
			high-temperature superconductors},\ }\href
	{https://doi.org/https://doi.org/10.1038/nphys2438} {\bibfield  {journal}
		{\bibinfo  {journal} {Nat. Phys.}\ }\textbf {\bibinfo {volume} {8}},\
		\bibinfo {pages} {709} (\bibinfo {year} {2012})}\BibitemShut {NoStop}%
	\bibitem [{\citenamefont {Alvarez}(2009)}]{Alvarez2009}%
	\BibitemOpen
	\bibfield  {author} {\bibinfo {author} {\bibfnamefont {G.}~\bibnamefont
			{Alvarez}},\ }\bibfield  {title} {\bibinfo {title} {The density matrix
			renormalization group for strongly correlated electron systems: A generic
			implementation},\ }\href {https://doi.org/10.1016/j.cpc.2009.02.016}
	{\bibfield  {journal} {\bibinfo  {journal} {Comp. Phys. Comms.}\ }\textbf
		{\bibinfo {volume} {180}},\ \bibinfo {pages} {1572} (\bibinfo {year}
		{2009})}\BibitemShut {NoStop}%
	\bibitem [{Sup()}]{Supplemental}%
	\BibitemOpen
	\href@noop {} {}\bibinfo {note} {See Supplemental Material.}\BibitemShut
	{Stop}%
	\bibitem [{\citenamefont {Dolfi}\ \emph {et~al.}(2015)\citenamefont {Dolfi},
		\citenamefont {Bauer}, \citenamefont {Keller},\ and\ \citenamefont
		{Troyer}}]{PhysRevB.92.195139}%
	\BibitemOpen
	\bibfield  {author} {\bibinfo {author} {\bibfnamefont {M.}~\bibnamefont
			{Dolfi}}, \bibinfo {author} {\bibfnamefont {B.}~\bibnamefont {Bauer}},
		\bibinfo {author} {\bibfnamefont {S.}~\bibnamefont {Keller}},\ and\ \bibinfo
		{author} {\bibfnamefont {M.}~\bibnamefont {Troyer}},\ }\bibfield  {title}
	{\bibinfo {title} {Pair correlations in doped {Hubbard} ladders},\ }\href
	{https://doi.org/10.1103/PhysRevB.92.195139} {\bibfield  {journal} {\bibinfo
			{journal} {Phys. Rev. B}\ }\textbf {\bibinfo {volume} {92}},\ \bibinfo
		{pages} {195139} (\bibinfo {year} {2015})}\BibitemShut {NoStop}%
	\bibitem [{\citenamefont {Kennedy}(1990)}]{Kennedy_1990}%
	\BibitemOpen
	\bibfield  {author} {\bibinfo {author} {\bibfnamefont {T.}~\bibnamefont
			{Kennedy}},\ }\bibfield  {title} {\bibinfo {title} {Exact diagonalisations of
			open spin-1 chains},\ }\href {https://doi.org/10.1088/0953-8984/2/26/010}
	{\bibfield  {journal} {\bibinfo  {journal} {J. Phys.: Condens. Matter}\
		}\textbf {\bibinfo {volume} {2}},\ \bibinfo {pages} {5737} (\bibinfo {year}
		{1990})}\BibitemShut {NoStop}%
	\bibitem [{\citenamefont {Qin}\ \emph {et~al.}(1995)\citenamefont {Qin},
		\citenamefont {Ng},\ and\ \citenamefont {Su}}]{PhysRevB.52.12844}%
	\BibitemOpen
	\bibfield  {author} {\bibinfo {author} {\bibfnamefont {S.}~\bibnamefont
			{Qin}}, \bibinfo {author} {\bibfnamefont {T.-K.}\ \bibnamefont {Ng}},\ and\
		\bibinfo {author} {\bibfnamefont {Z.-B.}\ \bibnamefont {Su}},\ }\bibfield
	{title} {\bibinfo {title} {Edge states in open antiferromagnetic {Heisenberg}
			chains},\ }\href {https://doi.org/10.1103/PhysRevB.52.12844} {\bibfield
		{journal} {\bibinfo  {journal} {Phys. Rev. B}\ }\textbf {\bibinfo {volume}
			{52}},\ \bibinfo {pages} {12844} (\bibinfo {year} {1995})}\BibitemShut
	{NoStop}%
	\bibitem [{\citenamefont {Calabrese}\ and\ \citenamefont
		{Cardy}(2004)}]{Calabrese2004}%
	\BibitemOpen
	\bibfield  {author} {\bibinfo {author} {\bibfnamefont {P.}~\bibnamefont
			{Calabrese}}\ and\ \bibinfo {author} {\bibfnamefont {J.}~\bibnamefont
			{Cardy}},\ }\bibfield  {title} {\bibinfo {title} {Entanglement entropy and
			quantum field theory},\ }\href
	{https://doi.org/10.1088/1742-5468/2004/06/p06002} {\bibfield  {journal}
		{\bibinfo  {journal} {J. Stat. Mech.}\ }\textbf {\bibinfo {volume} {2004}},\
		\bibinfo {pages} {P06002} (\bibinfo {year} {2004})}\BibitemShut {NoStop}%
	\bibitem [{\citenamefont {Gannot}\ \emph {et~al.}(2020)\citenamefont {Gannot},
		\citenamefont {Jiang},\ and\ \citenamefont {Kivelson}}]{PhysRevB.102.115136}%
	\BibitemOpen
	\bibfield  {author} {\bibinfo {author} {\bibfnamefont {Y.}~\bibnamefont
			{Gannot}}, \bibinfo {author} {\bibfnamefont {Y.-F.}\ \bibnamefont {Jiang}},\
		and\ \bibinfo {author} {\bibfnamefont {S.~A.}\ \bibnamefont {Kivelson}},\
	}\bibfield  {title} {\bibinfo {title} {Hubbard ladders at small $u$
			revisited},\ }\href {https://doi.org/10.1103/PhysRevB.102.115136} {\bibfield
		{journal} {\bibinfo  {journal} {Phys. Rev. B}\ }\textbf {\bibinfo {volume}
			{102}},\ \bibinfo {pages} {115136} (\bibinfo {year} {2020})}\BibitemShut
	{NoStop}%
	\bibitem [{\citenamefont {Okamoto}\ and\ \citenamefont
		{Millis}(2012)}]{PhysRevB.85.115406}%
	\BibitemOpen
	\bibfield  {author} {\bibinfo {author} {\bibfnamefont {J.-i.}\ \bibnamefont
			{Okamoto}}\ and\ \bibinfo {author} {\bibfnamefont {A.~J.}\ \bibnamefont
			{Millis}},\ }\bibfield  {title} {\bibinfo {title} {One-dimensional physics in
			transition-metal nanowires: Renormalization group and bosonization
			analysis},\ }\href {https://doi.org/10.1103/PhysRevB.85.115406} {\bibfield
		{journal} {\bibinfo  {journal} {Phys. Rev. B}\ }\textbf {\bibinfo {volume}
			{85}},\ \bibinfo {pages} {115406} (\bibinfo {year} {2012})}\BibitemShut
	{NoStop}%
	\bibitem [{\citenamefont {K\"onig}\ \emph {et~al.}(2018)\citenamefont
		{K\"onig}, \citenamefont {Tsvelik},\ and\ \citenamefont
		{Coleman}}]{PhysRevB.98.184517}%
	\BibitemOpen
	\bibfield  {author} {\bibinfo {author} {\bibfnamefont {E.~J.}\ \bibnamefont
			{K\"onig}}, \bibinfo {author} {\bibfnamefont {A.~M.}\ \bibnamefont
			{Tsvelik}},\ and\ \bibinfo {author} {\bibfnamefont {P.}~\bibnamefont
			{Coleman}},\ }\bibfield  {title} {\bibinfo {title} {Renormalization group
			analysis for the quasi-one-dimensional superconductor
			{${\mathrm{BaFe}}_{2}{\mathrm{S}}_{3}$}},\ }\href
	{https://doi.org/10.1103/PhysRevB.98.184517} {\bibfield  {journal} {\bibinfo
			{journal} {Phys. Rev. B}\ }\textbf {\bibinfo {volume} {98}},\ \bibinfo
		{pages} {184517} (\bibinfo {year} {2018})}\BibitemShut {NoStop}%
	\bibitem [{\citenamefont {Schulz}(1996)}]{PhysRevB.53.R2959}%
	\BibitemOpen
	\bibfield  {author} {\bibinfo {author} {\bibfnamefont {H.~J.}\ \bibnamefont
			{Schulz}},\ }\bibfield  {title} {\bibinfo {title} {Phases of two coupled
			{Luttinger} liquids},\ }\href {https://doi.org/10.1103/PhysRevB.53.R2959}
	{\bibfield  {journal} {\bibinfo  {journal} {Phys. Rev. B}\ }\textbf {\bibinfo
			{volume} {53}},\ \bibinfo {pages} {R2959} (\bibinfo {year}
		{1996})}\BibitemShut {NoStop}%
	\bibitem [{\citenamefont {Noack}\ \emph {et~al.}(1996)\citenamefont {Noack},
		\citenamefont {White},\ and\ \citenamefont {Scalapino}}]{NOACK1996281}%
	\BibitemOpen
	\bibfield  {author} {\bibinfo {author} {\bibfnamefont {R.~M.}\ \bibnamefont
			{Noack}}, \bibinfo {author} {\bibfnamefont {S.}~\bibnamefont {White}},\ and\
		\bibinfo {author} {\bibfnamefont {D.}~\bibnamefont {Scalapino}},\ }\bibfield
	{title} {\bibinfo {title} {The ground state of the two-leg {Hubbard} ladder a
			density-matrix renormalization group study},\ }\href
	{https://doi.org/https://doi.org/10.1016/S0921-4534(96)00515-1} {\bibfield
		{journal} {\bibinfo  {journal} {Physica C: Supercond.}\ }\textbf {\bibinfo
			{volume} {270}},\ \bibinfo {pages} {281} (\bibinfo {year}
		{1996})}\BibitemShut {NoStop}%
	\bibitem [{\citenamefont {Noack}\ \emph {et~al.}(1997)\citenamefont {Noack},
		\citenamefont {Bulut}, \citenamefont {Scalapino},\ and\ \citenamefont
		{Zacher}}]{PhysRevB.56.7162}%
	\BibitemOpen
	\bibfield  {author} {\bibinfo {author} {\bibfnamefont {R.~M.}\ \bibnamefont
			{Noack}}, \bibinfo {author} {\bibfnamefont {N.}~\bibnamefont {Bulut}},
		\bibinfo {author} {\bibfnamefont {D.~J.}\ \bibnamefont {Scalapino}},\ and\
		\bibinfo {author} {\bibfnamefont {M.~G.}\ \bibnamefont {Zacher}},\ }\bibfield
	{title} {\bibinfo {title} {Enhanced ${d}_{{x}^{2}\ensuremath{-}{y}^{2}}$
			pairing correlations in the two-leg {Hubbard} ladder},\ }\href
	{https://doi.org/10.1103/PhysRevB.56.7162} {\bibfield  {journal} {\bibinfo
			{journal} {Phys. Rev. B}\ }\textbf {\bibinfo {volume} {56}},\ \bibinfo
		{pages} {7162} (\bibinfo {year} {1997})}\BibitemShut {NoStop}%
	\bibitem [{\citenamefont {Shen}\ \emph {et~al.}(2023)\citenamefont {Shen},
		\citenamefont {Zhang},\ and\ \citenamefont {Qin}}]{Shen2023}%
	\BibitemOpen
	\bibfield  {author} {\bibinfo {author} {\bibfnamefont {Y.}~\bibnamefont
			{Shen}}, \bibinfo {author} {\bibfnamefont {G.-M.}\ \bibnamefont {Zhang}},\
		and\ \bibinfo {author} {\bibfnamefont {M.}~\bibnamefont {Qin}},\ }\bibfield
	{title} {\bibinfo {title} {Reexamining doped two-legged {Hubbard} ladders},\
	}\href {https://doi.org/10.1103/PhysRevB.108.165113} {\bibfield  {journal}
		{\bibinfo  {journal} {Phys. Rev. B}\ }\textbf {\bibinfo {volume} {108}},\
		\bibinfo {pages} {165113} (\bibinfo {year} {2023})}\BibitemShut {NoStop}%
	\bibitem [{\citenamefont {White}\ \emph {et~al.}(2002)\citenamefont {White},
		\citenamefont {Affleck},\ and\ \citenamefont
		{Scalapino}}]{PhysRevB.65.165122}%
	\BibitemOpen
	\bibfield  {author} {\bibinfo {author} {\bibfnamefont {S.~R.}\ \bibnamefont
			{White}}, \bibinfo {author} {\bibfnamefont {I.}~\bibnamefont {Affleck}},\
		and\ \bibinfo {author} {\bibfnamefont {D.~J.}\ \bibnamefont {Scalapino}},\
	}\bibfield  {title} {\bibinfo {title} {Friedel oscillations and charge
			density waves in chains and ladders},\ }\href
	{https://doi.org/10.1103/PhysRevB.65.165122} {\bibfield  {journal} {\bibinfo
			{journal} {Phys. Rev. B}\ }\textbf {\bibinfo {volume} {65}},\ \bibinfo
		{pages} {165122} (\bibinfo {year} {2002})}\BibitemShut {NoStop}%
	\bibitem [{Note1()}]{Note1}%
	\BibitemOpen
	\bibinfo {note} {\protect \leavevmode {\protect Our preliminary
			results on short chains with fixed $J_H/U=0.25$ suggest that i) the binding
			and pair-pair correlations are robust to the introduction of interorbital
			hopping, although the optimal $U/W$ ratio changes, ii) the binding and
			pair-pair correlations survive up to at least 10\% detuning of the diagonal
			intraorbital hopping, and iii) the physics is sensitive to crystal field
			effects. Indeed, a crystal field raising the energy of one orbital by $0.03W$
			already inhibits binding in $L=16$ chains. Thus deviations from a diagonal
			nearest-neighbor hopping matrix may be acceptable, but care should be taken
			to find materials with minimal crystal field splitting.}}\BibitemShut {Stop}%
	\bibitem [{\citenamefont {Xu}\ \emph {et~al.}(1996)\citenamefont {Xu},
		\citenamefont {DiTusa}, \citenamefont {Ito}, \citenamefont {Oka},
		\citenamefont {Takagi}, \citenamefont {Broholm},\ and\ \citenamefont
		{Aeppli}}]{PhysRevB.54.R6827}%
	\BibitemOpen
	\bibfield  {author} {\bibinfo {author} {\bibfnamefont {G.}~\bibnamefont
			{Xu}}, \bibinfo {author} {\bibfnamefont {J.~F.}\ \bibnamefont {DiTusa}},
		\bibinfo {author} {\bibfnamefont {T.}~\bibnamefont {Ito}}, \bibinfo {author}
		{\bibfnamefont {K.}~\bibnamefont {Oka}}, \bibinfo {author} {\bibfnamefont
			{H.}~\bibnamefont {Takagi}}, \bibinfo {author} {\bibfnamefont
			{C.}~\bibnamefont {Broholm}},\ and\ \bibinfo {author} {\bibfnamefont
			{G.}~\bibnamefont {Aeppli}},\ }\bibfield  {title} {\bibinfo {title}
		{{${\mathrm{Y}}_{2}$BaNi${\mathrm{O}}_{5}$}: A nearly ideal realization of
			the {$S=1$ Heisenberg} chain with antiferromagnetic interactions},\ }\href
	{https://doi.org/10.1103/PhysRevB.54.R6827} {\bibfield  {journal} {\bibinfo
			{journal} {Phys. Rev. B}\ }\textbf {\bibinfo {volume} {54}},\ \bibinfo
		{pages} {R6827} (\bibinfo {year} {1996})}\BibitemShut {NoStop}%
	\bibitem [{\citenamefont {Zhang}\ \emph
		{et~al.}(2022{\natexlab{a}})\citenamefont {Zhang}, \citenamefont {Lin},
		\citenamefont {Moreo},\ and\ \citenamefont {Dagotto}}]{10.1063/5.0079570}%
	\BibitemOpen
	\bibfield  {author} {\bibinfo {author} {\bibfnamefont {Y.}~\bibnamefont
			{Zhang}}, \bibinfo {author} {\bibfnamefont {L.-F.}\ \bibnamefont {Lin}},
		\bibinfo {author} {\bibfnamefont {A.}~\bibnamefont {Moreo}},\ and\ \bibinfo
		{author} {\bibfnamefont {E.}~\bibnamefont {Dagotto}},\ }\bibfield  {title}
	{\bibinfo {title} {{Electronic and magnetic properties of
				quasi-one-dimensional osmium halide OsCl4}},\ }\href
	{https://doi.org/10.1063/5.0079570} {\bibfield  {journal} {\bibinfo
			{journal} {Appl. Phys. Lett.}\ }\textbf {\bibinfo {volume} {120}},\ \bibinfo
		{pages} {023101} (\bibinfo {year} {2022}{\natexlab{a}})}\BibitemShut
	{NoStop}%
	\bibitem [{\citenamefont {Feng}\ \emph {et~al.}(2020)\citenamefont {Feng},
		\citenamefont {Patel}, \citenamefont {Kim}, \citenamefont {Han},\ and\
		\citenamefont {Trivedi}}]{PhysRevB.101.155112}%
	\BibitemOpen
	\bibfield  {author} {\bibinfo {author} {\bibfnamefont {S.}~\bibnamefont
			{Feng}}, \bibinfo {author} {\bibfnamefont {N.~D.}\ \bibnamefont {Patel}},
		\bibinfo {author} {\bibfnamefont {P.}~\bibnamefont {Kim}}, \bibinfo {author}
		{\bibfnamefont {J.~H.}\ \bibnamefont {Han}},\ and\ \bibinfo {author}
		{\bibfnamefont {N.}~\bibnamefont {Trivedi}},\ }\bibfield  {title} {\bibinfo
		{title} {Magnetic phase transitions in quantum spin-orbital liquids},\ }\href
	{https://doi.org/10.1103/PhysRevB.101.155112} {\bibfield  {journal} {\bibinfo
			{journal} {Phys. Rev. B}\ }\textbf {\bibinfo {volume} {101}},\ \bibinfo
		{pages} {155112} (\bibinfo {year} {2020})}\BibitemShut {NoStop}%
	\bibitem [{\citenamefont {Feng}\ \emph {et~al.}(2022)\citenamefont {Feng},
		\citenamefont {Alvarez},\ and\ \citenamefont
		{Trivedi}}]{PhysRevB.105.014435}%
	\BibitemOpen
	\bibfield  {author} {\bibinfo {author} {\bibfnamefont {S.}~\bibnamefont
			{Feng}}, \bibinfo {author} {\bibfnamefont {G.}~\bibnamefont {Alvarez}},\ and\
		\bibinfo {author} {\bibfnamefont {N.}~\bibnamefont {Trivedi}},\ }\bibfield
	{title} {\bibinfo {title} {Gapless to gapless phase transitions in quantum
			spin chains},\ }\href {https://doi.org/10.1103/PhysRevB.105.014435}
	{\bibfield  {journal} {\bibinfo  {journal} {Phys. Rev. B}\ }\textbf {\bibinfo
			{volume} {105}},\ \bibinfo {pages} {014435} (\bibinfo {year}
		{2022})}\BibitemShut {NoStop}%
	\bibitem [{\citenamefont {Zhang}\ \emph
		{et~al.}(2022{\natexlab{b}})\citenamefont {Zhang}, \citenamefont {Lin},
		\citenamefont {Moreo}, \citenamefont {Maier}, \citenamefont {Alvarez},\ and\
		\citenamefont {Dagotto}}]{PhysRevB.105.174410}%
	\BibitemOpen
	\bibfield  {author} {\bibinfo {author} {\bibfnamefont {Y.}~\bibnamefont
			{Zhang}}, \bibinfo {author} {\bibfnamefont {L.-F.}\ \bibnamefont {Lin}},
		\bibinfo {author} {\bibfnamefont {A.}~\bibnamefont {Moreo}}, \bibinfo
		{author} {\bibfnamefont {T.~A.}\ \bibnamefont {Maier}}, \bibinfo {author}
		{\bibfnamefont {G.}~\bibnamefont {Alvarez}},\ and\ \bibinfo {author}
		{\bibfnamefont {E.}~\bibnamefont {Dagotto}},\ }\bibfield  {title} {\bibinfo
		{title} {Strongly anisotropic electronic and magnetic structures in oxide
			dichlorides {${\mathrm{RuOCl}}_{2}$} and {${\mathrm{OsOCl}}_{2}$}},\ }\href
	{https://doi.org/10.1103/PhysRevB.105.174410} {\bibfield  {journal} {\bibinfo
			{journal} {Phys. Rev. B}\ }\textbf {\bibinfo {volume} {105}},\ \bibinfo
		{pages} {174410} (\bibinfo {year} {2022}{\natexlab{b}})}\BibitemShut
	{NoStop}%
	\bibitem [{\citenamefont {McGuire}(2020)}]{10.1063/5.0023729}%
	\BibitemOpen
	\bibfield  {author} {\bibinfo {author} {\bibfnamefont {M.~A.}\ \bibnamefont
			{McGuire}},\ }\bibfield  {title} {\bibinfo {title} {{Cleavable magnetic
				materials from van der Waals layered transition metal halides and
				chalcogenides}},\ }\href {https://doi.org/10.1063/5.0023729} {\bibfield
		{journal} {\bibinfo  {journal} {J. Appl. Phys.}\ }\textbf {\bibinfo {volume}
			{128}},\ \bibinfo {pages} {110901} (\bibinfo {year} {2020})}\BibitemShut
	{NoStop}%
	\bibitem [{\citenamefont {Laurell}\ \emph {et~al.}(2024)\citenamefont
		{Laurell}, \citenamefont {Herbrych}, \citenamefont {Alvarez},\ and\
		\citenamefont {Dagotto}}]{data}%
	\BibitemOpen
	\bibfield  {author} {\bibinfo {author} {\bibfnamefont {P.}~\bibnamefont
			{Laurell}}, \bibinfo {author} {\bibfnamefont {J.}~\bibnamefont {Herbrych}},
		\bibinfo {author} {\bibfnamefont {G.}~\bibnamefont {Alvarez}},\ and\ \bibinfo
		{author} {\bibfnamefont {E.}~\bibnamefont {Dagotto}},\ }\href
	{https://doi.org/10.5281/zenodo.12702044} {\bibinfo {title} {Data for
			"{Luther-Emery} liquid and dominant singlet superconductivity in the
			hole-doped {Haldane} spin-1 chain"}},\ \bibinfo {howpublished} {Zenodo}
	(\bibinfo {year} {2024})\BibitemShut {NoStop}%
\end{thebibliography}

\begin{thebibliography}{1}%
	\makeatletter
	\providecommand \@ifxundefined [1]{%
		\@ifx{#1\undefined}
	}%
	\providecommand \@ifnum [1]{%
		\ifnum #1\expandafter \@firstoftwo
		\else \expandafter \@secondoftwo
		\fi
	}%
	\providecommand \@ifx [1]{%
		\ifx #1\expandafter \@firstoftwo
		\else \expandafter \@secondoftwo
		\fi
	}%
	\providecommand \natexlab [1]{#1}%
	\providecommand \enquote  [1]{``#1''}%
	\providecommand \bibnamefont  [1]{#1}%
	\providecommand \bibfnamefont [1]{#1}%
	\providecommand \citenamefont [1]{#1}%
	\providecommand \href@noop [0]{\@secondoftwo}%
	\providecommand \href [0]{\begingroup \@sanitize@url \@href}%
	\providecommand \@href[1]{\@@startlink{#1}\@@href}%
	\providecommand \@@href[1]{\endgroup#1\@@endlink}%
	\providecommand \@sanitize@url [0]{\catcode `\\12\catcode `\$12\catcode
		`\&12\catcode `\#12\catcode `\^12\catcode `\_12\catcode `\%12\relax}%
	\providecommand \@@startlink[1]{}%
	\providecommand \@@endlink[0]{}%
	\providecommand \url  [0]{\begingroup\@sanitize@url \@url }%
	\providecommand \@url [1]{\endgroup\@href {#1}{\urlprefix }}%
	\providecommand \urlprefix  [0]{URL }%
	\providecommand \Eprint [0]{\href }%
	\providecommand \doibase [0]{https://doi.org/}%
	\providecommand \selectlanguage [0]{\@gobble}%
	\providecommand \bibinfo  [0]{\@secondoftwo}%
	\providecommand \bibfield  [0]{\@secondoftwo}%
	\providecommand \translation [1]{[#1]}%
	\providecommand \BibitemOpen [0]{}%
	\providecommand \bibitemStop [0]{}%
	\providecommand \bibitemNoStop [0]{.\EOS\space}%
	\providecommand \EOS [0]{\spacefactor3000\relax}%
	\providecommand \BibitemShut  [1]{\csname bibitem#1\endcsname}%
	\let\auto@bib@innerbib\@empty
	%</preamble>
	\bibitem [{\citenamefont {Patel}\ \emph {et~al.}(2017)\citenamefont {Patel},
		\citenamefont {Nocera}, \citenamefont {Alvarez}, \citenamefont {Moreo},\ and\
		\citenamefont {Dagotto}}]{PhysRevB.96.024520}%
	\BibitemOpen
	\bibfield  {author} {\bibinfo {author} {\bibfnamefont {N.~D.}\ \bibnamefont
			{Patel}}, \bibinfo {author} {\bibfnamefont {A.}~\bibnamefont {Nocera}},
		\bibinfo {author} {\bibfnamefont {G.}~\bibnamefont {Alvarez}}, \bibinfo
		{author} {\bibfnamefont {A.}~\bibnamefont {Moreo}},\ and\ \bibinfo {author}
		{\bibfnamefont {E.}~\bibnamefont {Dagotto}},\ }\bibfield  {title} {\bibinfo
		{title} {Pairing tendencies in a two-orbital {Hubbard} model in one
			dimension},\ }\href {https://doi.org/10.1103/PhysRevB.96.024520} {\bibfield
		{journal} {\bibinfo  {journal} {Phys. Rev. B}\ }\textbf {\bibinfo {volume}
			{96}},\ \bibinfo {pages} {024520} (\bibinfo {year} {2017})}\BibitemShut
	{NoStop}%
\end{thebibliography}
\end{document}